\documentclass[final,3p,times,onecolumn,authoryear]{elsarticle}
\usepackage{graphicx}
\usepackage{amssymb,amsmath,color}
\usepackage{subfigure}

\usepackage{epstopdf}
\usepackage[ruled]{algorithm2e}

\usepackage{array,booktabs}
\newcolumntype{L}{@{}>{\kern\tabcolsep}l<{\kern\tabcolsep}}
\usepackage{colortbl}
\usepackage{hyperref}
\usepackage{color}
\usepackage{xcolor}
\definecolor{dark-red}{rgb}{0.4,0.15,0.15}
\definecolor{dark-blue}{rgb}{0.15,0.15,0.4}
\definecolor{medium-blue}{rgb}{0,0,0.5}
\definecolor{bananamania}{rgb}{0.98, 0.91, 0.71}
\definecolor{}{rgb}{1.0, 0.97, 0.91}

\hypersetup{
  colorlinks   = true, 
  urlcolor     = blue, 
  linkcolor    = blue, 
  citecolor   = red 
}

\definecolor{darkred}{rgb}{0.4,0.15,0.15}
\definecolor{darkblue}{rgb}{0.15,0.15,0.4}
\definecolor{medium-blue}{rgb}{0,0,0.5}
\definecolor{darkgreen}{rgb}{0.0, 0.5, 0.0}
\definecolor{aqua}{rgb}{0.0, 1.0, 1.0}

\journal{Journal of Non-Newtonian Fluid Mechanics}

\begin{document}

\begin{frontmatter}

\title{A nonlinear dynamical system approach for the yielding behaviour of a viscoplastic fluid}
\date{today}

\author[RSMM]{Raazesh Sainudiin}
\author[RSMM]{Miguel Moyers-Gonzalez}
\address[RSMM]{School of Mathematics and Statistics, Private Bag 4800, University of Canterbury, Christchurch 8041, New Zealand}
\author[TB]{Teodor Burghelea \corref{cor1}}
\cortext[cor1]{Corresponding Author:  \ead{Teodor.Burghelea@univ-nantes.fr}, Tel.: + (33) - (0)2 40 68 31 85; Fax: + (33) - (0)2 40 68 31 41}


\address[TB]{Universit\'e de Nantes, Nantes Atlantique Universit\'es, CNRS, Laboratoire de Thermocin\'etique de Nantes, UMR 6607, La Chantrerie, Rue Christian Pauc, B.P. 50609, F-44306 Nantes Cedex 3, France}

\begin{abstract}

A nonlinear dynamical system model that approximates a microscopic Gibbs field model  for the yielding of a viscoplastic material subjected to varying external stress recently reported in \cite{Sainudiin15} is presented. The predictions of the model are in a fair agreement with the microscopic simulations and in a very good agreement with the micro-structural semi-empirical model reported in \cite{solidfluid}. 
With only two internal parameters, the nonlinear dynamical system model captures several key features of the solid-fluid transition observed in experiments: the effect of the interactions between microscopic constituents on the yield point, the abruptness of solid-fluid transition and  the emergence of a \textit{hysteresis} of the micro-structural states upon increasing/decreasing external forcing.The scaling behaviour of the magnitude of the hysteresis with the degree of the steadiness of the flow is consistent with previous experimental observations. 

\end{abstract}


\end{frontmatter}

\clearpage
\tableofcontents
\hypersetup{linkcolor=red}
\clearpage
\listoffigures
\hypersetup{linkcolor=red}
\clearpage

\section{Introduction}\label{S:Introduction}

A broad class of materials exhibit a dual response when subjected to an external tress. For low applied stresses they behave as solids (loosely speaking they may deform but they do not flow) but, if the stress  exceeds a critical threshold generally referred to as the \textit{``yield stress''}, they behave as fluids (typically non-Newtonian) and a macroscopic flow is observed. This distinct class of materials has been termed as \textit{``yield stress fluids''}.   During the past several decades one could notice an  increasing level of interest of both theoreticians and experimentalists in yield stress. This has a two-fold motivation. From a practical standpoint, such materials have found a significant number of  applications in several industries (which include food, cosmetical, pharmaceutical, oil field engineering, etc.) and they are encountered in daily life in various forms such as food pastes, hair gels and emulsions, cement, mud etc.. More recently, hydrogels which exhibit a yield stress have found a number of future promising applications including  targeted drug delivery \cite{jeong,Qiu2001321}, contact lenses, noninvasive intervertebral disc repair \cite{hou} and tissue engineering \cite{beck}.

From a fundamental standpoint, yield stress materials continue triggering intensive debates and posing difficult challenges to both theoreticians and experimentalists from various communities: soft matter physics,  rheology, physical chemistry and applied mathematics. The progress in understanding the flow behaviour of yield stress materials made the object of several review papers \cite{bogerreview,Coussot201431,ianreview,dennreview}. 
The best known debate concerning the yield stress materials is undoubtedly that related to the very existence of a "true" yield stress behaviour  \cite{barnes1,barnes2}. During the past two decades, however, a number of innovative improvements in the rheometric equipment made possible measurements of torques as small as $0.1 nNm$  and rates of deformation as small as $10^{-7}~s^{-1}$). Such accurate rheological measurements proved unequivocally the existence of a true yielding behaviour \cite{moller1, solidfluid, dennyieldstress, dennrheoacta}. The physics of the yielding process itself on the other hand remains elusive. The macroscopic response of yield stress fluids subjected to an external stress $\sigma$  has been classically described by the Herschel-Bulkely model \cite{originalhb,H-B}:

\begin{equation}
\sigma = \sigma_y + K \dot{\gamma}^{N}
\label{eq:hb}
\end{equation}

Here $\sigma_y$ is the yield stress, $\dot{\gamma}$ is the rate of shear, i.e., the rate at which the material is being deformed,  $\sigma$ is the macroscopically applied stress (the external forcing parameter), $K$ is a so-called consistency parameter that sets the viscosity scale in the flowing state and $N$ is the power law index which characterises the degree of shear thinning of the viscosity beyond the yield point. 

In spite of its wide use by rheologists, fluid dynamicists and engineers, the Herschel-Bulkley model (and its regularised variants, e.g. Papanastasiou \cite{Papanastasiou1987}) is in fact applicable only for a limited number of yield stress materials, sufficiently far from the solid-fluid transition, i.e.~when $\sigma > \sigma_y$, and in the conditions of a steady state forcing, i.e.~ when a constant external stress $\sigma$ is applied over a long period of time.  

Thixotropy, which may be loosely understood \footnote{A universal consensus on the definition of thixotropy has not yet been reached, Chapter 9.2 in Ref. \cite{handbook} provides as many as $7$ alternative definitions.} as a time dependence of the rheological parameters which results from a competition between destruction and rejuvenation of the soft material units subjected to stress, is considered to be a major reason for the departure from this simple yielding picture \cite{moller}. It has been recently suggested that a number of difficulties concerning the yielding behaviour of yield stress materials could be solved if a clear distinction between thixotropic and non-thixotropic yield stress fluids is made \cite{Moller5139}. 

However, it has been shown recently that a clear departure from the Herschel-Bulkley behaviour can be observed even for non-thixotropic yield stress fluids such as the Carbopol gels particularly during either controlled stressed rheological tests \cite{solidfluid,miguelstab,thermo,divoux3,divoux4} or hydrodynamically \textit{``simple''} flow problems such as a creeping motion of a spherical object \cite{sedimentation}, the unsteady laminar pipe flow \cite{unsteady} or the emergence of the Rayleigh-B\'{e}nard convection, \cite{teoconvection} in a yield stress fluid heated from below.  

To overcome these difficulties, several phenomenological macroscopic models have been proposed \cite{dullaert,quemada1,quemada3,quemada4,avalanchecoussot, avalanchecoussot1, coussotthixotropy,solidfluid,paulo1,paulo2,garethyielding,garethlaos,randymodel} which have a general form \footnote{For simplicity, only a scalar form is given but they can be written in a tensorial form as well.}: 

\begin{equation}
\frac{d \bar {a}  (t)}{dt} =F\left[ \bar {a} (t), \sigma(t), C_1, C_2, ..., C_m  \right]
\label{eq:structuralmodels}
\end{equation}

The particular feature of these models is that they describe the temporal evolution of a microstructural parameter $\bar a (t)$ as a function of the applied stress and a number of parameters $C_1, ..., C_m$. Part of these parameters describe the kinetics of the destruction/restructuration of the material and are difficult (or impossible!) to measure. The rest of the parameters are measurable via adequate macroscopic rheological tests (flow ramps, oscillatory measurements, creep/relaxation tests etc.).  Thus, as the applied stress is increased, $\overline{a}(t)$ varies smoothly from $1$ (the entire volume of material is in a solid state) to $0$ (the entire volume of material is in a fluid state) and the combined solid and fluid rheological responses are accounted for accordingly into a constitutive relation. Finally, the problem is reduced to a system of coupled equation which can be solved either analytically if the structural evolution equation is sufficiently simple (see for example the $\lambda$ model proposed by Coussot in \cite{avalanchecoussot1}, \cite{coussotthixotropy}) or numerically.  

Such approaches have several clear indisputable advantages and have contributed significantly to our current understanding of yield stress materials:

\begin{enumerate}

\item{ As opposed to the Bingham and the Herschel-Bulkley models which predict an abrupt and discontinuous solid-fluid transition when $\overline{a}$ jumps from $1$ to $0$ at a well-defined value of the applied stress which coincides with the yield stress $\sigma=\sigma_y$, such approaches which directly account for the evolution of $\overline{a}(t)$ are able to predict a smooth (gradual) solid-fluid transition which is often observed in experiments \cite{solidfluid}}.
 
 \item{From the point of view of a rheologist, these models are quite versatile, as with a minimal readjustment of the parameters  they can model various types of tests: flow ramps \cite{solidfluid}, small amplitude oscillations (SAOS) \cite{solidfluid}, large amplitude oscillatory flows (LAOS) \cite{solidfluid,garethlaos,randymodel} and creep/relaxation flows.} 
  
\item{ They can flexibly model the irreversibility of the deformation states when the applied stresses are increased/decreased around the solid-fluid transition. Moreover, they can quantitatively describe the experimentally observed rheological hysteresis and the dependence of its magnitude on the degree of steadiness of the forcing \cite{solidfluid} as reflected by the area of the hysteresis of the dependence $\dot \gamma = \sigma \left( \dot \gamma \right)$, on the degree of the steadiness of the external forcing (i.e.~how fast is the external stress varied about the solid-fluid transition).}

\item{From a practical standpoint, they are relatively easy to implement. }
\end{enumerate}

Though able to model sufficiently complex rheological data (ranging from controlled stress/strain unsteady flow ramps, creep tests and oscillatory tests in a wide range of frequencies and amplitudes), such phenomenological macroscopic models do have a number of limitations:

\begin{enumerate}
\item{As the functional dependence $F$ in Eq.~\ref{eq:structuralmodels} is generally chosen on an intuitive basis rather derived from first principles, these models can teach little about the microscopic scale physics of the yielding process.}
\item {They typically involve a rather large number of parameters some of which are not directly and easily measurable and can be obtained only by fitting the experimental data.}
\item{ Second, such models are not inherently validated from a thermodynamical standpoint. The second law of thermodynamics is not guaranteed to be held and such a validation is not always straightforward as it requires the derivation of a thermodynamic potential \cite{picardfluidity,manero,Hong20081779}}. 

\end{enumerate}     
     
Bearing in mind that the yield stress behaviour originates from the presence of a \textit{``soft''} microstructure which can only  sustain a finite local stress prior to its breakdown, an alternative way of assessing the dynamics of the yielding process is to focus on the evolution of the micro-structural soft material units as the external stress is gradually increased past the solid-fluid transition and next to assess the macroscopic scale behaviour from the perspective of statistical mechanics. 
A thermodynamic approach for the deformation of a physical gel has been recently proposed by An and coworkers \cite{solis}.  By using a mean field approach, they construct a free energy functional and describe the microscopic scale dynamics of  the gel network as a function of the applied stress in terms of the monomer volume fraction and an internal connectivity tensor characterising the gel network. Peshkov and his coworkers have employed the irreversible mechanics and thermodynamics of two-phase continua to describe the yielding process but a comparison with the experimental data seems to require future developments \cite{Peshkov2014}.
 \citet{debruyn} has modelled the restricted diffusion of small tracer particles in heterogeneous media by performing Monte Carlo simulations in a site-percolation model and his results partially agreed with the experimental observations  \cite{debruyn1,debruyn2}. 

More recently, we have proposed  in  \cite{Sainudiin15} a microscopic picture of yielding inspired from the Ising model of magnetisation of a ferromagnet \cite{ising,stanley}.
The model was built on an analogy between the local agglomerative interactions in terms of assembly/disassembly of neighbouring microscopic constituents in a yield stress material subjected to an external stress and the local ferromagnetic interactions in terms of spin up ($+1$) / spin down ($-1$) of neighbouring particles in a microscopic ferromagnetic system subjected to an external magnetic field. First, our approach is fundamentally probabilistic and formalises Gibbs fields as time-homogeneous and time-inhomogeneous Markov chains over the state space of all microscopic configurations and thus it is thermodynamically validated. Second, the model has solely two parameters. Though able to capture several key physical features of the solid-fluid transition, from a practical perspective the applicability of this model to describe rheological measurements and real flow problems is somewhat limited. First, the microscopic constituents are assumed fixed over a lattice while the external stress is varied and thus there exists no direct of equivalent of the rate of deformation $\dot \gamma$. Second, as it is a statistical model, its implementation is not trivial and its usage is time consuming. The aim of the current contribution is to derive an approximation of the model in the form of a classical structural approach similar to Eq.~\ref{eq:structuralmodels} which, together with the appropriate constitutive equation could ultimately describe rheological measurements. 

The paper is organised as follows. 
In Sec.~\ref{S:Model} we provide a brief description of the microscopic Gibbs field approach introduced in   \cite{Sainudiin15} with a particular emphasis on its main predictions. 
A differential equation approximating the expected solid fraction of the material in the Gibbs model is derived and analysed in Sec.~\ref{S:ApproxModel}.  
The results of the simulations according to the microscopic model and the expected trajectories from the approximating differential equation are presented in Sec.~\ref{S:Results}.  
The paper concludes in Sec.~\ref{S:Discussion} with a discussion of the main findings, their impact and their possible implications and extensions.

\section{The microscopic Gibbs field model}\label{S:Model}

First we present a summary of the microscopic Gibbs Field model by Sainudiin et. al. \cite{Sainudiin15}. We model an idealised yield stress material or viscoplastic fluid as a network of particles in an appropriate solvent that are capable of assembling by ``forming bonds'' or disassembling by "breaking bonds'' with their neighbours when an external stress $\sigma$ is applied. As already mentioned in the introduction, this approach is inspired by the Ising model of ferromagnetism and its advantage is that, once formulated, it can fully benefit from the already developed tools of Statistical Physics \cite{stanley,landaust}.

We investigate the model when the network of particles is the regular graph given by the toroidal two-dimensional square lattice. 
Let $x_s \in \Lambda=\{0,1\}$ denote the phase at site $s$.  
Phase $0$ corresponds to being {\it yielded} or {\it ungelled} and phase $1$ corresponds to being {\it unyielded} or {\it gelled}.  
The phase at a site directly affects its {\it connectability} with its neighbuoring sites.  
We assume that only two gelled sites can be connected with one another.  

We consider the following Gibbs potential over the two types of cliques:

\begin{equation}
V_{\{s\}}(x) = (\sigma - \alpha) x_s = 
\begin{cases}
0 & \text{ if } x_s=0 \\
\sigma - \alpha & \text{ if } x_s=1,
\end{cases}
\end{equation}
and
\begin{equation}
V_{\langle s,r\rangle}(x) = -\beta x_s x_r = 
\begin{cases}
0 & \text{ if } (x_s,x_r)=(0,0) \\
0 & \text{ if } (x_s,x_r)=(1,0) \\
0 & \text{ if } (x_s,x_r)=(0,1) \\
-\beta & \text{ if } (x_s,x_r)=(1,1) ,
\end{cases}
\end{equation}

where $\{s\}$ is the singleton clique, $\langle s,r\rangle$ is the doubleton clique with $r\in N_s$ (the set of four nearest neighbouring sites of a given site $s$), 
$\sigma \geq 0$ is the external stress applied, $\alpha \geq 0$ is the {\it site-specific threshold}, and $\beta \in (-\infty,\infty)$ is the {\it interaction constant} between neighbouring sites. It is important to note that the "stress" $\sigma$ has actually the dimensions of an energy transferred to the lattice which is an obvious consequence of the fact that the sites in the lattice are fixed and there is no equivalent of a deformation $\gamma$ of the lattice. 

Using the Gibbs potentials defined above one can write the associated energy of a site configuration as:

\begin{eqnarray}
\mathcal{E}(x) 
&=& \sum_{C} V_C(x) \nonumber  \\
&=&  \sum_{s \in \mathbb{S}_n} V_{\{s\}}(x) + \sum_{\langle s,r \rangle \in \mathbb{E}_n} V_{\langle s,r \rangle}(x) \nonumber \\
&=& \left(- \beta \sum_{\langle s,r \rangle \in \mathbb{E}_n} x_s x_r + (\sigma-\alpha) \sum_{s \in \mathbb{S}_n} x_s \right) \enspace .
\end{eqnarray}  

As expected, the energy function above is very similar to the Ising Hamiltonian, \cite{ising,stanley}. Here the external stress $\sigma$ is the analogue of an external magnetic field and the interaction parameter parameter $\beta$ plays the role of the coupling between neighbouring magnetic spins. 

The probability distribution of interest on the site configuration space $\mathbb{X}_n$ is then
\begin{equation}
\pi(x) = \frac{1}{Z_{kT}} \exp \left( -\frac{1}{kT} \mathcal{E}(x) \right) 
\end{equation}

where $Z_{kT}$ is the normalizing constant or partition function
\begin{equation}
Z_{kT} = \sum_{x \in \mathbb{X}_n} \exp \left( -\frac{1}{kT} \mathcal{E}(x) \right)  .
\end{equation}
\subsection{Local Specification}

Let the number of neighbours of site $s$ that are in phase $1$ be $x_{N_s} := \sum_{r \in N_s}x_r$.  
Then, $\mathcal{E}_s(x)$, the {\it local energy} at site $s$ of configuration $x$, is obtained by summing the Gibbs potential $V_C(x)$ over all $C \ni s$, i.e., over cliques $C$ containing site $s$, as follows
\begin{eqnarray}
\mathcal{E}_s(x) &=& \sum_{C \ni s} V_C(x) = V_{\{s\}}(x) + \sum_{r \in N_s} V_{\langle s,r \rangle}(x) \nonumber \\
&=& (\sigma - \alpha)x_s - \beta \sum_{r \in N_s} x_sx_r \nonumber \\
&=& x_s  \left( (\sigma - \alpha) - \beta \sum_{r \in N_s} x_r \right) \nonumber \\
&=& x_s  \left( (\sigma - \alpha) - \beta x_{N_s} \right) \enspace .
\end{eqnarray}
Let $(\lambda,x(\mathbb{S} \setminus s))$ denote the configuration that 
is in phase $\lambda$ at $s$ and identical to $x$ everywhere else.  
Then the {\it local specification} is
\begin{eqnarray}\label{E:LocSpec}
\pi_s(x) 
&=& \frac{\exp(-\frac{1}{kT} \mathcal{E}_s(x))}{\sum_{\lambda \in \Lambda}\exp(-\frac{1}{kT} \mathcal{E}_s(\lambda,x(\mathbb{S}\setminus s)))} \notag \\ 
&=&
\begin{cases}
\frac{\theta}{1+\theta} & \text{ if } x_s=0\\
\frac{1}{1+\theta} & \text{ if } x_s=1
\end{cases}\enspace ,
\end{eqnarray}
where
\begin{equation}\label{E:theta}
\theta = \theta(s,\alpha,\beta,\sigma) = \exp\left(-\frac{1}{k T} \left( \beta x_{N_s}-(\sigma-\alpha) \right) \right) \enspace .
\end{equation}
In this work we focus on the effect of varying external stress $\sigma$ at a constant ambient temperature, and therefore without loss of generality, we take $kT=1$ and work with $\pi(x) = Z_1^{-1}\exp(-\mathcal{E}(x))$.

\subsection{Markov chain on configuration space}

We can think of the microscopic Gibbs field model as an $\mathbb{X}_n$-valued Markov chain $\{X(m)\}_{m = 0}^{\infty}$, where $X(m)=\left(X_s(m), s \in \mathbb{S}_n\right)$ and $X_s(m) \in \Lambda$, in discrete time $m \in \mathbb{Z}_+ := \{0,1,2,\ldots\}$.  
Let the initial condition, $X(0)=x(0)$, be given by the initial distribution $\delta_{x(0)}$ over $\mathbb{X}_n$ that is entirely concentrated at state $x(0)$.  
Then the conditional probability of the Markov chain at time-step $m$, given that it starts at time $0$ in state $x(0)$, is
\begin{equation}\label{E:PXmGivenx0}
\Pr \left\{ \, X(m) \, | \, X(0)=x(0) \, \right\} = \delta_{x(0)} \left(P_{\alpha,\beta,\sigma}\right)^m \enspace,
\end{equation}
where, the $|\mathbb{X}_n| \times |\mathbb{X}_n|$ transition probability matrix $P_{\alpha,\beta,\sigma}$ over any pair of configurations $(x,x') \in \mathbb{X}_n \times \mathbb{X}_n$ is
\begin{equation}\label{E:P_abs}
P_{\alpha,\beta,\sigma} (x,x')
= 
\begin{cases}
\frac{1}{n^2} \frac{1}{1+\theta} & \text{ if } || x-x' || = 1, 0 = x_s \neq x'_s = 1 \\
\frac{1}{n^2} \frac{\theta}{1+\theta} & \text{ if } || x-x' || = 1, 1 = x_s \neq x'_s = 0 \\
\frac{1}{n^2} \frac{1}{1+\theta} & \text{ if } || x-x' || = 0, 1 = x_s = x'_s = 1 \\
\frac{1}{n^2} \frac{\theta}{1+\theta} & \text{ if } || x-x' || = 0, 0 = x_s = x'_s = 0 \\
0 & \text { otherwise}\enspace.
\end{cases}
\end{equation}
and $\theta=\theta(s,\alpha,\beta,\sigma)$, is indeed a function of the site $s$ and the three parameters: $\alpha$, $\beta$ and $\sigma$.  
By $|| x-x' || = 1$ we mean that the configurations $x$ and $x'$ differ at exactly site $s$, i.e., $x_s \neq x'_s$.  
Similarly, by $|| x-x' || = 0$ we mean that the two configurations are identical, i.e., $x=x'$ or $x_s=x'_s$ at every site $s \in \mathbb{S}_n$.  
We can think of our Markov chain evolving according to the following probabilistic rules based on \eqref{E:LocSpec} and \eqref{E:theta}: 
\begin{itemize}
\item given the current configuration $x$, we first choose one of the $n^2$ sites in $\mathbb{S}_n$ uniformly at random with probability $n^{-2}$, 
\item denote this chosen site by $s$ and let the number of bondable neighbors of $s$ be $i=N_s(x) \in \{0,1,2,3,4\}$, and 
\item finally change the phase at $s$ to $1$, i.e., set $x_s=1$ with probability 
\begin{equation}\label{E:psOfsigma}
p_i := (1+\theta)^{-1} = (1+\theta(s,\alpha,\beta,\sigma))^{-1} = 1/(1+e^{(\sigma-\alpha-i\beta)})
\end{equation}
and set $x_s=0$ with probability $1-p_i$.
\end{itemize}

\begin{figure}[htbp]
\begin{center}
\centering
\makebox{
{\includegraphics[width=0.65\textwidth]{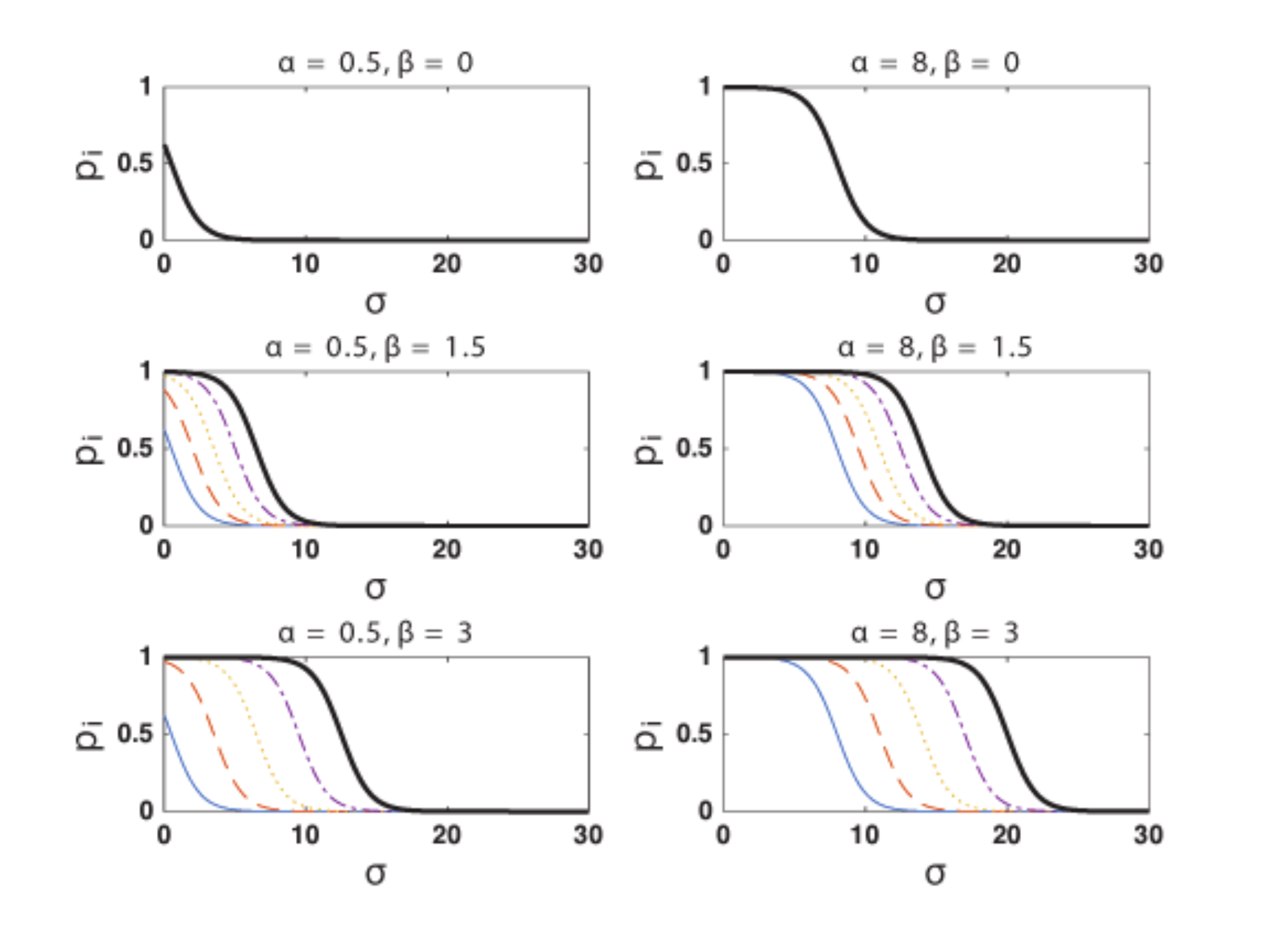}}
}
\end{center}
\caption{Plots of $p_i$, the probability that site $s$ with $i=x_{N_s}$ neighbors in phase $1$, is also in phase $1$, as a function of external stress $\sigma$ for different values of $\alpha$, $\beta$. $p_0$ (full line), $p_1$ (--), $p_2$ (..), $p_3$ (-.-) and $p_4$ (bold line). 
\label{F:ProbOf1s}}
\end{figure}

We emphasise the dependence of $p_i$ on the parameters $\alpha$, $\beta$ and $\sigma$ by $p_i(\alpha,\beta,\sigma)$.  
This is plotted in Fig.~\ref{F:ProbOf1s} for different parameter values.  From the plots it is clear that $\alpha$ is a location parameter while $\beta$ controls the scale of the relative difference between $p_i$'s.  

\subsection{Sufficient Configuration Statistics for Energy}\label{subsec:energyconfiguration}

Two informative singleton clique statistics of a configuration $x(m)$ at time $m$ are the number and fraction of gelled sites, given respectively by:
\[
a(x) := \sum_{s \in \mathbb{S}_n} x_s \quad \text{and} \quad \overline{a}(x) := |\mathbb{S}_n|^{-1} a(x)  = \frac{a(x)}{n^2}\enspace .
\]

Similarly, two informative doubleton clique statistics of a configuration $x$ are the number and fraction of connected pairs of neighboring sites, given respectively by:
\begin{align*}
b(x) 
&:= \sum_{\langle s,t \rangle \in \mathbb{E}_n} y_{\langle s,r \rangle} = \sum_{\langle s,r \rangle \in \mathbb{E}_n} x_r x_s 
\quad \text{and} \\
\overline{b}(x) 
&:= |\mathbb{E}_n|^{-1} b(x) = \frac{b(x)}{2n^2} 
\enspace .
\end{align*}

When the configuration is a function of time $m$ and given by $x(m)$, then the corresponding configuration statistics are also functions of time and are given by: $a(m)=a(x(m))$, $\overline{a}(m)  = \overline{a}(x(m))$, $b(m)=b(x(m))$ and $\overline{b}(m)  = \overline{b}(x(m))$.  
The energy of a configuration $x$ can be succinctly expressed in terms of $\overline{a}(x)$ and $\overline{b}(x)$ as
\[
\mathcal{E}(x) = -\beta b(x) + (\sigma - \alpha) a(x)  = - \beta 2n^2 \overline{b}(x) + (\sigma - \alpha) n^2 \overline{a}(x) \enspace ,
\]
and therefore
\begin{equation}\label{E:EnergyPropTo}
\mathcal{E}(x) \propto -2\beta \overline{b}(x) + (\sigma - \alpha) \overline{a}(x) = -2\beta \overline{b}(x) + \tilde{\sigma} \overline{a}(x)\enspace ,
\end{equation}
where $\beta \in (-\infty,\infty)$ and $\tilde{\sigma} = \sigma - \alpha \geq -\alpha$ for a given $\alpha \geq 0$.  
Since the energy of a configuration $x$, given $n$, only depends on its $\overline{a}(x)$ and $\overline{b}(x)$, we can easily visualize any sample path 
$\left( \, x(0),\ldots,x(m) \, \right) \in \mathbb{X}_n^{m+1}$ in configuration space that is outputted by either Algorithm~1 or Algorithm~2 presented in \cite{Sainudiin15} as the following sequence of $(m+1)$ ordered pairs in the unit square:
\[
\left( \, \left(\overline{a}(x(0)), \overline{b}(x(0))\right), \ldots , \left(\overline{a}(x(m)), \overline{b}(x(m))\right) \, \right) \in \left([0,1]^2\right)^{m+1} \enspace .
\] 

Finally, we reserve upper-case letters for random variables.  
Thus, $A(X)$, $\overline{A}(X)$, $B(X)$ and $\overline{B}(X)$ are the statistics of the random configuration $X$.  
And the notation naturally extends to $A(m)$, $\overline{A}(m)$, $B(m)$ and $\overline{B}(m)$ when $X(m)$ is a random configuration at time $m$. 
  
\subsection{The main predictions of the Gibbs field model for the yielding of a yield stress material}\label{gibbsmodelpredictions}

 The Gibbs field model has been tested by monitoring the evolution of the volume fraction of the solid microscopic constituents $\bar a (t)$ during increasing/decreasing linearly stepped stress ramp that mimics a rheological flow ramp, see \cite{solidfluid, miguelstab,thermo}.  In the context of the Gibbs field model, the closest equivalent of the characteristic forcing time $t_0$ or the time the stress is maintained constant during a controlled stress stepped ramped is the average number of hits per lattice site $h$ in the Gibbs algorithm.
 
 Corresponding to each step of the ramp the stress was kept constant during a time $t_0$ and the volume fraction of un-yielded lattice sites was obtained via the Gibbs algorithm detailed in the Appendix of Ref. \cite{Sainudiin15}. By varying the characteristic forcing time $t_0$ we could test the dependence of the microscopic yielding dynamics on the degree of steadiness of the external forcing and attempt a qualitative comparison with the experimental results referred to in Sec. \ref{S:Introduction}.
 
  In spite of its very limited number of parameters, we have shown that this model can capture several key features of the solid-fluid transition:
\begin{enumerate}

\item{In the limit of a steady state external forcing ($t_0$ very large), the solid-fluid transition is reversible upon increasing/decreasing applied stresses only if the interaction parameter $\beta$ does not exceed a critical threshold $\beta_c$, see Figs. 10 (a,b) in Ref. \cite{Sainudiin15}. Beyond this threshold a \textit{``genuine''} micro-structural hysteresis is observed even in the asymptotic limit of a steady state forcing, see Figs. 10 (c,d) in Ref. \cite{Sainudiin15}.}

\item{During unsteady flow ramps ($t_0$ finite) a micro-structural hysteresis is observed even in the absence of interaction ($\beta=0$). The dependence of the magnitude of the hysteresis on $t_0$ is non monotonic,  see Figs. 11 in Ref. \cite{Sainudiin15}. At large $t_0$ it scales as a power law $t_0^{-\xi}$ with $\xi$ decreasing as the interaction parameter $\beta$ increases which is qualitatively similar the experimentally observed rheological hysteresis behaviour \cite{solidfluid,miguelstab,thermo} }
For highly unsteady stress ramps (small $t_0$) the magnitude of the hysteresis follows a log-normal correlation which is equally consistent with the experimental scaling measured during rheological tests \cite{divoux4}.
\end{enumerate}
 
To conclude this part, we have been able to qualitatively describe several main features of the solid-fluid transition experimentally observed for yield stress materials subjected to an external stress using a Gibbs field statistical approach with only two internal parameters. This motivates us to derive in the following an approximate continuous version of this model which is similar in form to the classical micro-structural approaches generally described by Eq.~\ref{eq:structuralmodels} but remains thermodynamically validated.  
  
\section{An Approximating Nonlinear Dynamical System}\label{S:ApproxModel}

Here we derive a nonlinear first-order differential equation to asymptotically approximate $\mathbf{E}(\overline{A}(t))$, the expected fraction of sites in the solid phase, in continuous time $t$ that is measured in units of $n^2$ discrete time-steps as the number of sites $n^2 \to \infty$, under a fixed externally applied stress $\sigma$ and fixed rheological parameters $\alpha$ and $\beta$.

First consider the discrete-time Markov chain $\{X(m)\}_{m=0}^{\infty}$ of \eqref{E:PXmGivenx0} and \eqref{E:P_abs} and recall that 
$X(m)$ is the random site configuration of the chain at discrete time $m$ and 
$A(m)=\sum_{s}X_s(m)$ is the number of sites that are in phase $1$.  
We will derive the approximation first for the case when $\beta=0$ in \eqref{E:P_abs} and then for the general setting of $\beta \neq 0$.

\subsection{Non-interactive case with $\beta=0$}\label{subsec:noninteractive}
If $\beta=0$ then the probability of the phase in site $s$ at the next time-step is independent of the current configuration, i.e.,
\begin{multline*}
\Pr\left\{ X_s(m+1)=x_s(m+1) \, \middle\vert \, X(m)=x(m) \right\} \\
=\Pr\left\{ X_s(m+1)=x_s(m+1) \right\} \qquad \qquad \qquad \qquad\\
=
\begin{cases}
p = \left(1+e^{\sigma-\alpha}\right)^{-1} & \text{ if } x_s(m+1)=1\\
1-p = 1-\left(1+e^{\sigma-\alpha}\right)^{-1} & \text{ if } x_s(m+1)=0\\
0 & \text{ if } x_s(m+1) \notin \{0,1\} \enspace .
\end{cases}
\end{multline*}

Therefore, the probability that the total number of sites in phase $1$ increases by $1$ in one time-step is obtained by adding the probability of a transition from phase $0$ to phase $1$ over every uniformly chosen site $s$ as follows:
\begin{multline*}
\Pr \left\{ A(m+1)=a(m)+1 \, \middle\vert \, A(m)=a(m) \right\}\\
=
\sum_{s \in \mathbb{S}_n} \Pr \left\{ X_s(m+1)=1 , X_s(m)=0, S=s \, \middle\vert \, A(m)=a(m) \right\}\\
=
\sum_{s \in \mathbb{S}_n} \underset{p}{\underbrace{\Pr \left\{ X_s(m+1)=1  \, \middle\vert \, X_s(m)=0, S=s, A(m)=a(m) \right\}}} \\
\qquad \qquad \times \, \underset{(n^2-a(m))/n^2}{\underbrace{\Pr \left\{ X_s(m)=0  \, \middle\vert \, S=s, A(m)=a(m) \right\}}} \\
\qquad \qquad \times \, \underset{1/n^{2}}{\underbrace{\Pr \left\{ S=s \, \middle\vert \, A(m)=a(m) \right\}}} \\
=
\sum_{s \in \mathbb{S}_n} p \left(1-\frac{a(m)}{n^2}\right)  \frac{1}{n^2}  = p \left(1-\overline{a}(m)\right) \enspace .
\end{multline*}
Dividing both sides of the equality that defines the above event by $n^2$ we get
\begin{multline*}
\Pr \left\{ A(m+1)/n^2=a(m)/n^2+1/n^2 \, \middle\vert \, A(m)/n^2=a(m)/n^2 \right\} \\
= 
\Pr \left\{ \overline{A}(m+1)=\overline{a}(m)+1/n^2 \, \middle\vert \, \overline{A}(m)=\overline{a}(m) \right\} = p \left(1-\overline{a}(m)\right)\enspace .
\end{multline*}
By an analogous argument we can obtain the probabilities for the remaining two possibilities
\begin{multline*}
\Pr \left\{ \overline{A}(m+1)=\overline{a}(m)-1/n^2 \, \middle\vert \, \overline{A}(m)=\overline{a}(m) \right\} = (1-p) \overline{a}(m)\enspace ,\\
\Pr \left\{ \overline{A}(m+1)=\overline{a}(m) \, \middle\vert \, \overline{A}(m)=\overline{a}(m) \right\} = p\overline{a}(m) + (1-p) (1-\overline{a}(m)) \enspace .
\end{multline*}

Now we can define a continuous-time Markov chain $\{\overline{A}(t)\}_{t \geq 0}$ on the unit interval $[0,1]$ by a rescaling of the discrete-time Markov chain $\{\overline{A}(m)\}_{m = 0}^{\infty}$ and letting the number of sites $n^2 \to \infty$.  
These two Markov chains are notationally distinguished only by their time indices.
The rescaled time $t$ is $m$ in units of $n^2$, i.e., $m=\lfloor tn^2 \rfloor$ and $m+1=\lfloor(t+1/n^2)n^2\rfloor$.  
Then by taking $\Delta_t=O(1/n^2)$ and letting
$$\Delta_A = \overline{A}(t+\Delta_t)-\overline{a}(t)  = \overline{A}(\lfloor (t+\Delta_t)n^2 \rfloor)-\overline{a}(\lfloor tn^2 \rfloor) \enspace,$$ 
we get
\begin{multline}\label{E:LimitCondProbBeta0}
\Pr \left\{ \frac{\Delta_A}{\Delta_t}=\frac{\Delta_a}{\Delta_t} \, \middle\vert \, \overline{A}(t)=\overline{a}(t) \right\} \\
=
\begin{cases}
p \left(1-\overline{a}(t)\right) + O(\Delta_t) & \text{ if } \frac{\Delta_a}{\Delta_t}=1 \\
(1-p) \overline{a}(t) + O(\Delta_t) & \text{ if } \frac{\Delta_a}{\Delta_t}=-1 \\
p \overline{a}(t)+ (1-p)(1-\overline{a}(t))+O(\Delta_t) & \text{ if } \frac{\Delta_a}{\Delta_t}=0 \\
O(\Delta_t) & \text{ otherwise}\enspace .
\end{cases}
\end{multline}
Finally by considering the instantaneous rate of change of the expected fraction of sites in phase $1$
\[
\frac{d}{dt}\overline{\mathbf{a}}(t) := \lim_{\Delta_t \to 0} \mathbf{E} \left( \frac{\overline{A}(t+\Delta_t) - \overline{A}(t)}{\Delta_t} \, \middle\vert \, \overline{A}(t) \right)
\]
we get the limiting differential equation approximation as 
$$n^2 \to \infty, \quad \Delta_t \to 0, \quad \Delta_a \to 0 \enspace,$$ 
such that 
$$\Pr\{ \, \Delta_a/\Delta_t \in \{0,-1,+1\} \, \} \to 1$$ 
based on \eqref{E:LimitCondProbBeta0} as follows:
\[
\dot{\overline{\mathbf{a}}} = \frac{d}{dt}{\overline{\mathbf{a}}(t)} = p(1-\overline{\mathbf{a}}(t)) - (1-p) \overline{\mathbf{a}}(t) = p-\overline{\mathbf{a}}(t) \enspace ,
\]
or simply by
\begin{equation}\label{E:ODEAbarBeta0}
\dot{\overline{\mathbf{a}}} = p-\overline{\mathbf{a}} = (1+e^{\sigma-\alpha})^{-1}  - \overline{\mathbf{a}} \enspace .
\end{equation}

The simple relationship above is mathematically very similar to the so called ``lambda-model'' introduced in \cite{avalanchecoussot,avalanchecoussot1} with the remark that we consider the stress $\sigma$ as a forcing parameter rather than the rate of deformation.

Given the initial condition $\overline{\mathbf{a}}(0)=\overline{\mathbf{a}}_0$, the analytic solution is
\[
\overline{\mathbf{a}}(t) = p + (\overline{\mathbf{a}}_0-p)e^{-t} = (1+e^{\sigma-\alpha})^{-1} + (\overline{\mathbf{a}}_0 - (1+e^{\sigma-\alpha})^{-1}) e^{-t}
\]
with only one asymptotically stable fixed point 
\begin{equation}\label{E:FixedPointAbarBeta0}
\overline{\mathbf{a}}^* = p = (1+e^{\sigma-\alpha})^{-1} \enspace .
\end{equation}
Thus, $\overline{\mathbf{a}}(t)$ in the above differential equation is the expected fraction of sites in phase $1$ at time $t$ in the limit of an infinite toroidal square lattice with $|\mathbb{S}_n|=n^2\to \infty$ and a realization of the continuous time Markov chain $\{\overline{A}(t)\}_{t\geq 0}$ is $\overline{a}(t)$.

Since $\beta=0$, the probability of a site being in a given phase is independent of the phases of its neighboring sites.  
Thus, we can obtain $\overline{\mathbf{b}}(t)$, the expected fraction of bonds, by simply multiplying $\overline{\mathbf{a}}(t)$, the probability of finding a randomly chosen site in phase $1$, by itself, i.e.,
\begin{equation}\label{E:ODEBbarBeta0AndFixedPoint}
\overline{\mathbf{b}}(t) = \overline{\mathbf{a}}(t)^2 \text{ and } \overline{\mathbf{b}}^*=\left(\overline{\mathbf{a}}^*\right)^2\enspace .
\end{equation}

\subsection{Interactive case with $\beta \neq 0$}\label{S:binomApprox}

If $\beta \neq 0$ then the probability of site $s$ being in phase $1$ at time $m+1$ depends on the configuration of the neighboring sites of $s$ at time $m$ through $X_{N_s}(m)=\sum_{r \in N_s}X_r(m)$, the number of neighboring sites of $s$ in phase $1$ at time $m$.

\begin{multline*}
\Pr\left\{ X_s(m+1)=x_s(m+1) \, \middle\vert \, X(m)=x(m) \right\} \\
=\Pr\left\{ X_s(m+1)=x_s(m+1) \, \middle\vert \, X_{N_s}(m)=i \right\} \\
=
\begin{cases}
p_i = \left(1+e^{\sigma-\alpha-i\beta}\right)^{-1} & \text{ if } x_s(m+1)=1\\
1-p_i = 1-\left(1+e^{\sigma-\alpha-i\beta}\right)^{-1} & \text{ if } x_s(m+1)=0\\
0 & \text{ if } x_s(m+1) \notin \{0,1\} \enspace .
\end{cases}
\end{multline*}

Thus the probability that the phase changes from $0$ to $1$ in one time-step at site $s$ given that $a(m)$ is the total number of sites in phase $1$ at time $m$ is
\begin{multline*}
\Pr \left\{ X_s(m+1)=1 , X_s(m)=0 \, \middle\vert \, S=s, A(m)=a(m) \right\} \\
= \sum_{i=0}^{4} \Pr \left\{ X_s(m+1)=1 , X_{N_s}(m)=i , X_s(m)=0 \right.\\
\qquad \qquad \qquad \qquad \qquad \left. \, \middle\vert \, S=s, A(m)=a(m) \right\}\\
=\sum_{i=0}^{4}
\Pr \left\{ X_s(m+1)=1 \, \middle\vert \, X_{N_s}(m)=i ,\right. \qquad \qquad \qquad \qquad \\
\qquad \quad \underset{p_i}{\underbrace{\left. \qquad \qquad \qquad \qquad X_s(m)=0, S=s, A(m)=a(m) \right\}}} \\
\qquad \times \Pr \left\{ X_{N_s}(m)=i \, \middle\vert \, X_s(m)=0, S=s, A(m)=a(m) \right\}\\
\qquad \times \underset{(n^2-a(m))/n^2=1-\overline{a}(m)}{\underbrace{\Pr \left\{X_s(m)=0 \, \middle\vert \,  S=s, A(m)=a(m)\right\}}}\\
\end{multline*}

Since there are $4!/((4-i)!i!)$ distinct neighborhood configurations with $i$ of the four nearest neighbors of site $s$ in phase $1$, we can make the following binomial approximation for $\Pr \left\{ X_{N_s}(m)=i \, \middle\vert \, X_s(m)=0, S=s, A(m)=a(m) \right\}$ in the above expression and obtain
\begin{multline*}
\Pr \left\{ X_s(m+1)=1 , X_s(m)=0 \, \middle\vert \, S=s, A(m)=a(m) \right\} \\
=\sum_{i=0}^{4} p_i (1-\overline{a}(m)) \qquad \qquad \qquad \qquad \qquad \qquad \\
\qquad \qquad \times \Pr \left\{ X_{N_s}(m)=i \, \middle\vert \, X_s(m)=0, S=s, A(m)=a(m)\right\}\\
\approxeq \sum_{i=0}^{4} p_i \left(1-\overline{a}(m)\right) \, \binom{4}{i} \, \left(\overline{a}(m)\right)^i \, \left(1-\overline{a}(m)\right)^{4-i}\enspace .
\end{multline*}

Therefore, the probability that the total number of sites in phase $1$ increases by $1$ in one time-step is obtained by adding the probability of a transition from phase $0$ to phase $1$ over every uniformly chosen site $s$ as follows:
\begin{multline*}
\Pr \left\{ A(m+1)=a(m)+1 \, \middle\vert \, A(m)=a(m) \right\}\\
=
\sum_{s \in \mathbb{S}_n} \Pr \left\{ X_s(m+1)=1 , X_s(m)=0,  S=s \, \middle\vert \, A(m)=a(m) \right\}\\
=
\sum_{s \in \mathbb{S}_n} \Pr \left\{ X_s(m+1)=1, X_s(m)=0  \, \middle\vert \,  S=s, A(m)=a(m) \right\} \\
\qquad \qquad \times \, \underset{1/n^{2}}{\underbrace{\Pr \left\{ S=s \, \middle\vert \, A(m)=a(m) \right\}}} \\
\approxeq \sum_{s \in \mathbb{S}_n} \left( \sum_{i=0}^{4} p_i \left(1-\overline{a}(m)\right) \, \binom{4}{i} \, \left(\overline{a}(m)\right)^i \, \left(1-\overline{a}(m)\right)^{4-i} \right) \frac{1}{n^2}\\
= \left(1-\overline{a}(m)\right) \sum_{i=0}^{4} p_i \, \binom{4}{i} \, \left(\overline{a}(m)\right)^i \, \left(1-\overline{a}(m)\right)^{4-i} \enspace . 
\end{multline*}
Dividing both sides of the equality that defines the above event by $n^2$ we get
\begin{multline*}
\Pr \left\{ \overline{A}(m+1)=\overline{a}(m)+1/n^2 \, \middle\vert \, \overline{A}(m)=\overline{a}(m) \right\} \\
\approxeq 
\left(1-\overline{a}(m)\right) \sum_{i=0}^{4} p_i \, \binom{4}{i} \, \left(\overline{a}(m)\right)^i \, \left(1-\overline{a}(m)\right)^{4-i}\enspace .  
\end{multline*}
By an analogous argument we can obtain the probability that $\overline{A}(m+1)$ decreases by $1/n^2$ as
\begin{multline*}
\Pr \left\{ \overline{A}(m+1)=\overline{a}(m)-1/n^2 \, \middle\vert \, \overline{A}(m)=\overline{a}(m) \right\} \\
\approxeq 
\overline{a}(m) \sum_{i=0}^{4} \left( 1-p_i \right) \, \binom{4}{i} \, \left(\overline{a}(m)\right)^i \, \left(1-\overline{a}(m)\right)^{4-i}  \enspace .
\end{multline*}

Using the same limiting approximation in the previous Section we can obtain the following differential equation approximation for $\overline{\mathbf{a}}=\overline{\mathbf{a}}(t)$ 
\begin{eqnarray*} 
\dot{\overline{\mathbf{a}}} &=& \frac{d}{dt}\overline{\mathbf{a}}(t)\\
&=&(1-\overline{\mathbf{a}}) \left( p_0 \, (1-\overline{\mathbf{a}})^4 + p_1 \, 4 \overline{\mathbf{a}} (1-\overline{\mathbf{a}})^3 \right.\\
&~& \quad \left. + p_2 \, 6 \overline{\mathbf{a}}^2 (1-\overline{\mathbf{a}})^2 +  p_3 \, 4 \overline{\mathbf{a}}^3 (1-\overline{\mathbf{a}}) + p_4 \, \overline{\mathbf{a}}^4 \right)\\
&~&-\overline{\mathbf{a}} \left((1-p_0)\,(1-\overline{\mathbf{a}})^4 + (1-p_1)\, 4\overline{\mathbf{a}}(1-\overline{\mathbf{a}})^3 \right.\\
&~& \quad \left. + (1-p_2) \, 6 \overline{\mathbf{a}}^2 (1-\overline{\mathbf{a}})^2 + (1-p_3) \, 4 \overline{\mathbf{a}}^3 (1-\overline{\mathbf{a}}) \right.\\
&~& \quad \left. + (1-p_4) \, \overline{\mathbf{a}}^4) \right)\enspace .
\end{eqnarray*}
This simplifies after factoring and extracting coefficients of $\overline{\mathbf{a}}$ as follows:
\begin{eqnarray}\label{E:ODEAbarBetaGEQ0} 
\dot{\overline{\mathbf{a}}}(t) &=&
p_{0} -{\left(4 \, p_{0} - 4 \, p_{1} + 1\right)} \overline{\mathbf{a}} 
+ 6 \, {\left(p_{0} - 2 \, p_{1} + p_{2}\right)} \overline{\mathbf{a}}^{2} \notag \\
&~& -4 \, {\left(p_{0} - 3 \, p_{1} + 3 \, p_{2} - p_{3}\right)} \overline{\mathbf{a}}^{3} \notag \\
&~& + {\left(p_{0} - 4 \, p_{1} + 6 \, p_{2} - 4 \, p_{3} + p_{4}\right)} \overline{\mathbf{a}}^{4} \enspace .
\end{eqnarray}

We can understand \eqref{E:ODEAbarBetaGEQ0} directly as a quartic polynomial in $\overline{\mathbf{a}}$ whose coefficients are given by an alternating binomial series corresponding to the increase and decrease in $\overline{\mathbf{a}}$ based on a combinatorial averaging over the transition diagram of site configurations at the four nearest neighbors of a given site.  
Next we characterize the qualitative asymptotic dynamics of the above nonlinear differential equation which reduces to the differential equation \eqref{E:ODEAbarBeta0} if $\beta=0$ and thereby $p=p_0=p_1=p_2=p_3=p_4$.  

\begin{figure}[htbp]
\begin{center}
\centering
\makebox{
{\includegraphics[width=0.45\textwidth]{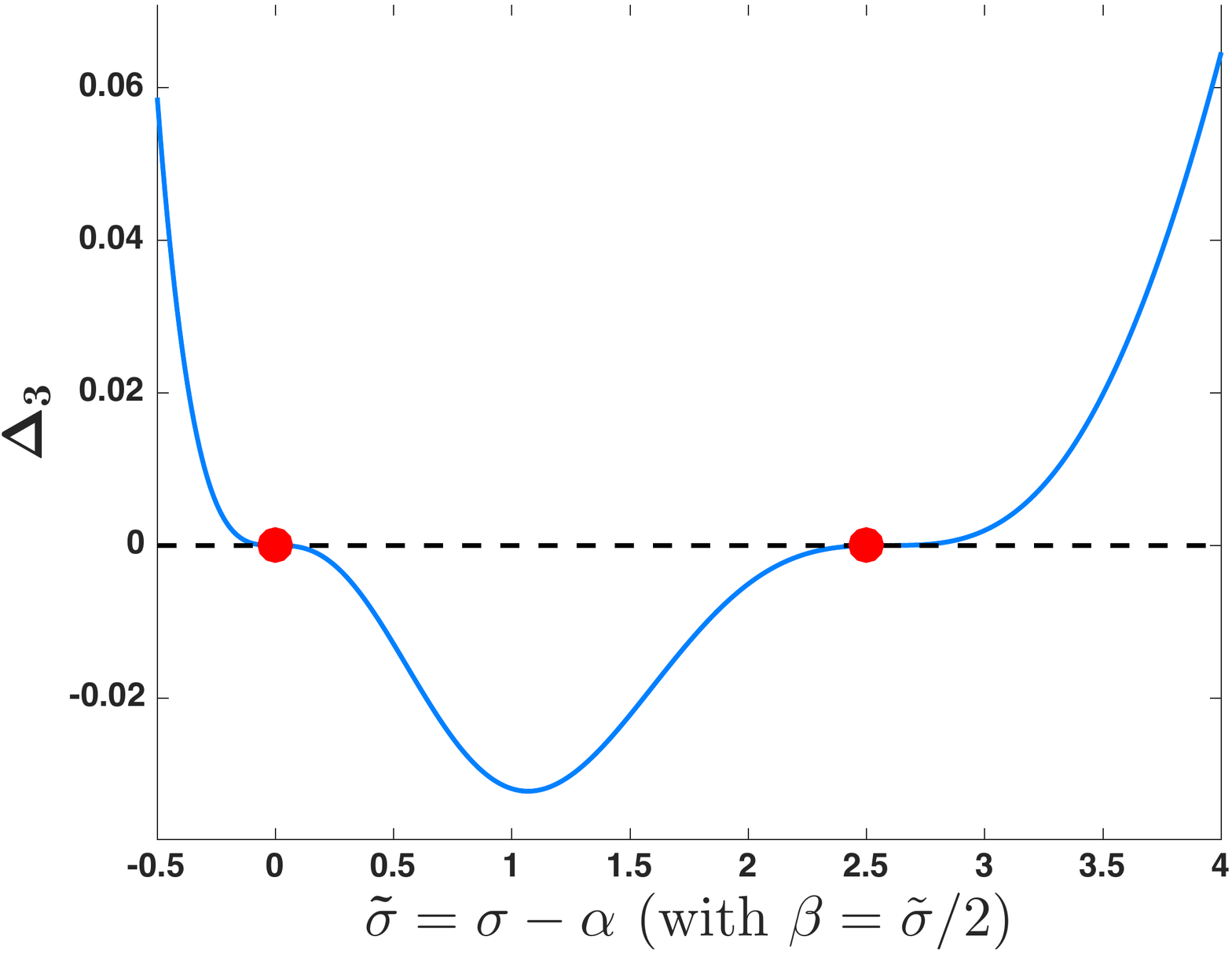}}
}
\end{center}
\caption{Discriminant $\Delta_3$ of the cubic function of $\overline{\mathbf{a}}$ along $\beta=\tilde{\sigma}/2$ as a function of $\tilde{\sigma}=\sigma-\alpha$.\label{F:Discriminant3}}
\end{figure}

If the externally applied stress $\sigma$ is beyond $\alpha$ by $2\beta$, i.e.
$$\tilde{\sigma}=\sigma-\alpha=2\beta \enspace ,$$
then the probability of being in phase $1$ or phase $0$ at a site that is surrounded by two neighbors in phase $1$ and the other two in phase $0$ is equal and exactly half: 
\[
p_2 = \frac{1}{1+\exp(\tilde{\sigma}-2(\tilde{\sigma}/2))}=\frac{1}{2} = 1-p_2 \enspace .
\]
If we study the system along $\tilde{\sigma}=2\beta$, the symmetric set of parameters, then
\[
p_1+p_3 = \frac{1}{1+e^{\tilde{\sigma}/2}}+\frac{1}{1+e^{-\tilde{\sigma}/2}}=1 \enspace ,
\]
and also
\[
p_0+p_4 = \frac{1}{1+e^{\tilde{\sigma}}}+\frac{1}{1+e^{-\tilde{\sigma}}}=1 \enspace .
\]
Thus, the coefficient of $\overline{\mathbf{a}}^4$ in \eqref{E:ODEAbarBetaGEQ0} vanishes when $\tilde{\sigma}=2\beta$:
\[
p_0 - 4 \, p_1 + 6 \, p_2 - 4 \, p_3 + p_4 = (p_0+p_4) -4(p_1+p_3)+6 p_2 = 0 \enspace.
\]
Therefore, along $\tilde{\sigma}=2\beta$ our \eqref{E:ODEAbarBetaGEQ0} is really just a cubic function of $\overline{\mathbf{a}}$ as opposed to a quartic.  
The discriminant of this cubic is
\[
\Delta_3(\tilde{\sigma},\beta) = 18 c_3 c_2 c_1 c_0 -4 c_2^3 c_0 + c_2^2 c_1^2 - 4 c_3 c_1^3 - 27 c_3^2 c_0^2 \enspace ,
\]
where $c_i=c_i(\tilde{\sigma},\beta)$ is the coefficient of $\overline{\mathbf{a}}^i$ in \eqref{E:ODEAbarBetaGEQ0},
and it takes negative values when $\tilde{\sigma} \in (0, 2.589145)$ (giving one real and two complex conjugate roots), 
takes positive values when $\tilde{\sigma}<0$ and $\tilde{\sigma}>2.589145$ (giving three distinct real roots) and 
takes $0$ when $\tilde{\sigma} \in \{0,2.589145\}$ (giving three multiple real roots) as shown in Fig.~\ref{F:Discriminant3}.  

\begin{figure}[htbp]
\begin{center}
\centering
\makebox{
{\includegraphics[width=0.5\textwidth]{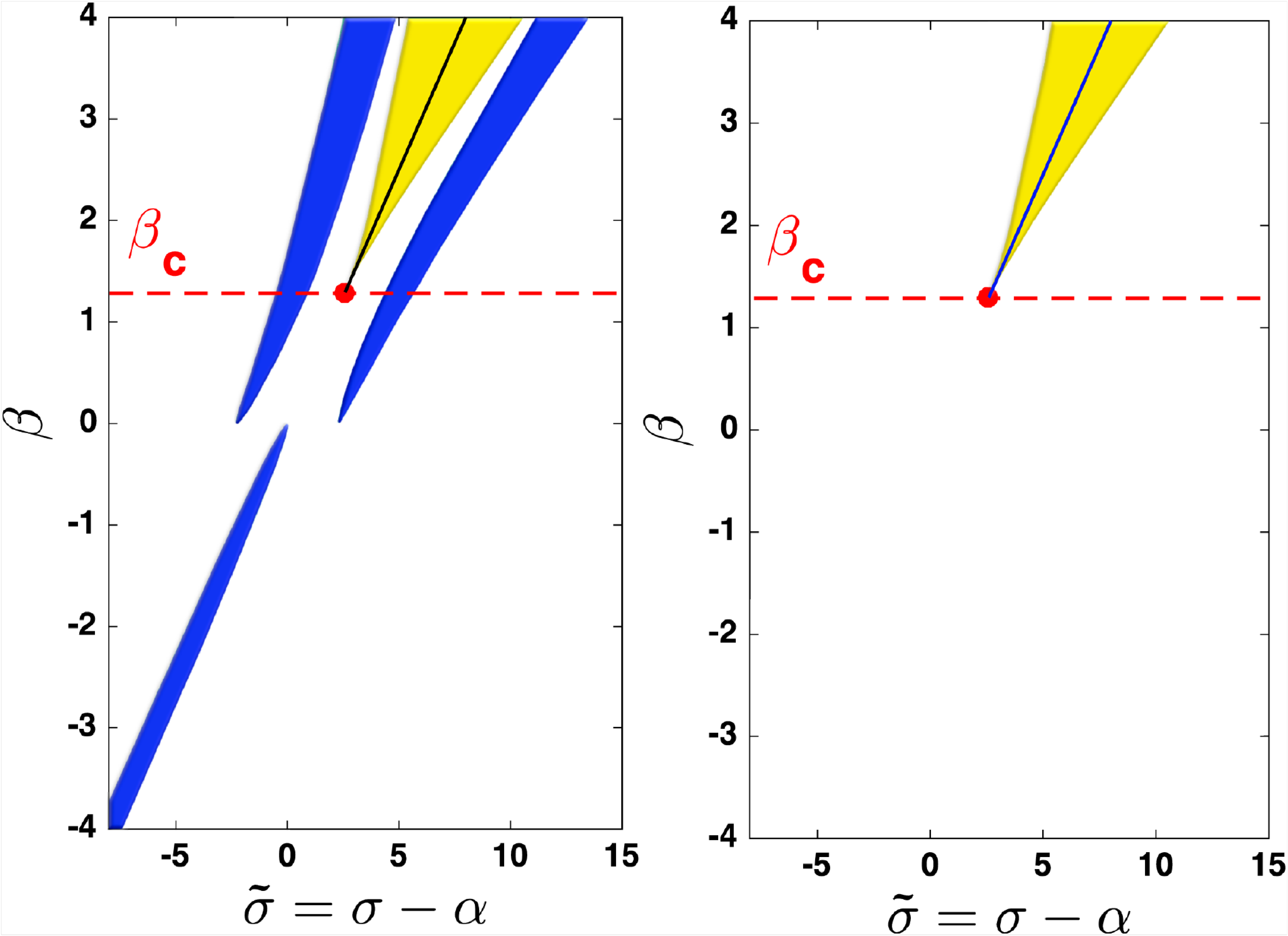}}
}
\end{center}
\caption{Four real roots of the quartic occur in the shaded regions (blue and yellow) over $\tilde{\sigma}=\sigma-\alpha$ and $\beta$ is shown in the left panel.  
The black line is $\beta=\tilde{\sigma}/2$ started at $(2.589145,1.2945725)$.  
The parameter space with only three distinct real roots in $[0,1]$ is shown in the right panel.
\label{F:Distinct4RealRoots}}
\end{figure}

A sign analysis of the discriminant of the quartic with sixteen terms: 
\begin{multline*}
\Delta_4(\tilde{\sigma},\beta) = 256 c_4^3 c_0^3 
   - 192 c_4^2 c_3 c_1 c_0^2 
   - 128 c_4^2 c_2^2 c_0^2\\
   + 144 c_4^2 c_2 c_1^2 c_0 
   - 27 c_4^2 c_1^4 
   + 144 c_4 c_3^2 c_2 c_0^2 
   - 6 c_4 c_3^2 c_1^2 c_0\\
   - 80 c_4 c_3 c_2^2 c_1 c_0 
   + 18 c_4 c_3 c_2 c_1^3 
   + 16 c_4 c_2^4 c_0
   - 4 c_4 c_2^3 c_1^2 \\
   - 27 c_3^4 c_0^2 
   + 18 c_3^3 c_2 c_1 c_0 
   - 4 c_3^3 c_1^3 
   - 4 c_3^2 c_2^3 c_0 
   + c_3^2 c_2^2 c_1^2
\end{multline*}
and the three associated polynomials:
\begin{eqnarray*}
D_4(\tilde{\sigma},\beta) &=& 64 c_4^3 c_0 
    - 16 c_4^2 c_2^2 
    + 16 c_4 c_3^2 c_2 \notag \\
    &~& - 16 c_4^2 c_3 c_1 
    - 3 c_3^4\\
\Delta_0(\tilde{\sigma},\beta) &=& 256 c_2^2 
    - 3 c_3 c_1 
    + 12 c_4 c_0\\
P_4(\tilde{\sigma},\beta) &=& 8 c_4 c_2 
    - 3 c_3^2 \enspace ,
\end{eqnarray*}
shows that the four real distinct roots occur inside the shaded region (blue and yellow regions) of the parameter space in the left panel of Fig.~\ref{F:Distinct4RealRoots} where $\Delta_4>0$, $P_4<0$ and $D_4<0$.  

In the left panel of Fig.~\ref{F:Distinct4RealRoots}, we present three different stability scenarios for the fixed points of equation (\ref{E:ODEAbarBetaGEQ0}) in the $(\tilde{\sigma},\beta)$ plane: (i) In the blue shaded region the right hand side of equation (\ref{E:ODEAbarBetaGEQ0}) has four real roots and only one of them is in $[0,1]$, this fixed point is stable. (ii) In the yellow region, starting at point $(2.589145,1.2945725)$, we have four distinct real roots with three of them in $[0,1]$. Only one of the three distinct real roots is an unstable fixed point while the other two roots are stable fixed points. This naturally corresponds to a family of pitch-fork bifurcations and the associated hysteresis depending on where the system is initialised from. (iii) The unshaded region in the left panel of Fig.~\ref{F:Distinct4RealRoots} corresponds to the parameter space where the quartic discriminant $\Delta_4$ is negative and thus implying the existence of two real roots (with one of them in $[0,1]$, stable fixed point) and two complex conjugate roots.

The real roots and their derivatives over each $(\tilde{\sigma},\beta)$ in a grid of parameter values from $[-8,12]\times [-4,4]$ were obtained through interval analytic methods using \cite{HofschusterK03}.

\begin{figure*}[htbp]
\begin{center}
\centering
\makebox{
{\includegraphics[width=0.5\textwidth, angle=270]{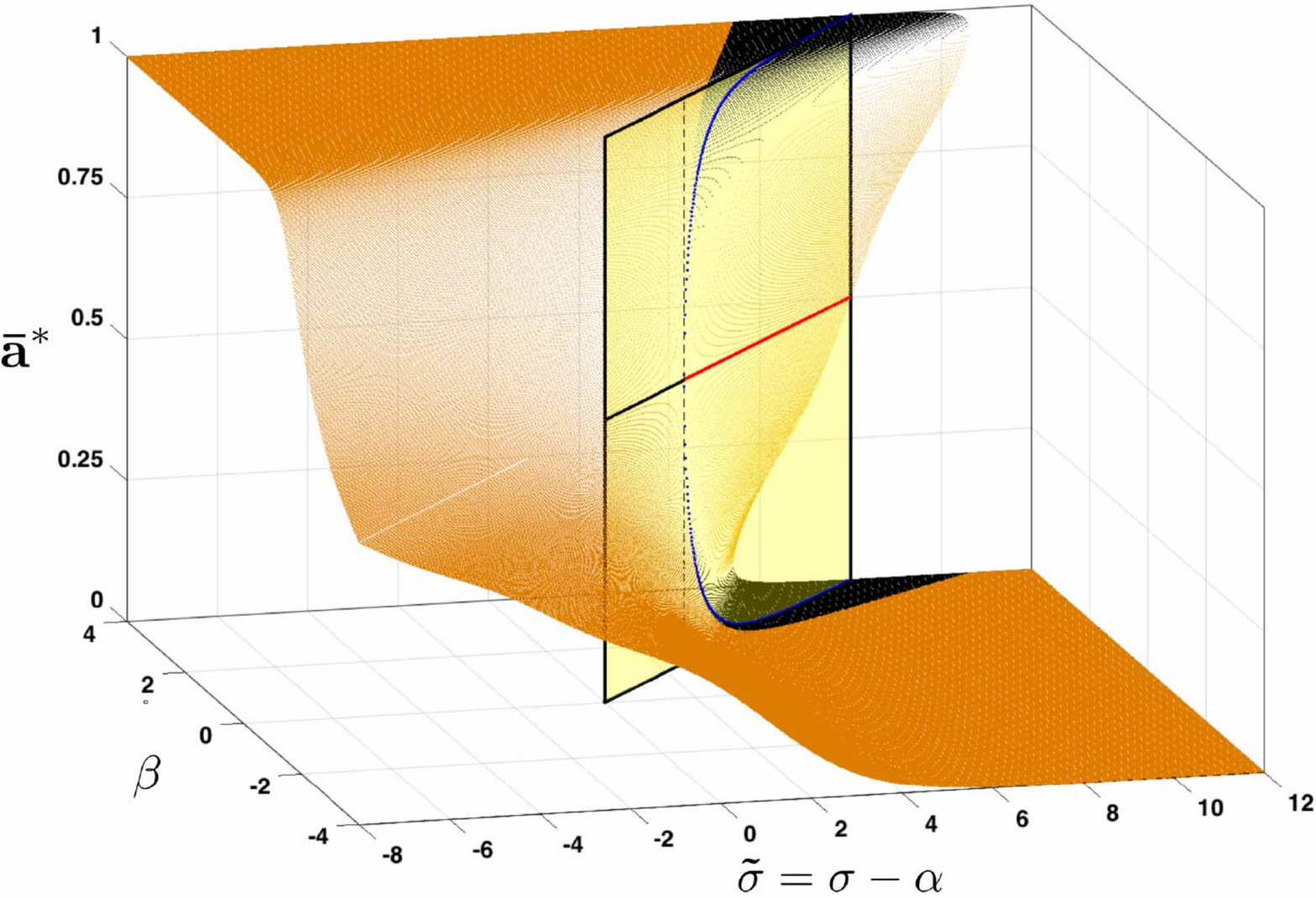}}
}
\end{center}
\caption{The fixed points $\overline{\mathbf{a}}^*$ as a set-valued function of the parameters $\tilde{\sigma}=\sigma-\alpha$ and $\beta$.  
The blue, black and azure points are the stable fixed points while the red and green points are the unstable fixed points of the system.  
There is a pitch-fork bifurcation along $\tilde{\sigma}=2\beta$ that starts at $(2.589145,1.2945725)$ where the fixed point at $0.5$ becomes unstable with two stable fixed points on either side.\label{F:Hyst}}
\end{figure*}

Figure~\ref{F:Hyst} shows the set of fixed points $\overline{\mathbf{a}}^*$ of the dynamical system as a function of $(\tilde{\sigma},\beta)$.  
The parameter space corresponding to the central shaded region 
of Fig.~\ref{F:Distinct4RealRoots} 
containing the line $\beta=\tilde{\sigma}/2$ 
is evident in Fig.~\ref{F:Hyst} with three fixed points in $[0,1]$.  
The pitch-fork bifurcations along the plane $\tilde{\sigma}=2\beta$ or $\beta=\tilde{\sigma}/2$ determined by the non-negative sign of the cubic discriminant of Fig.~\ref{F:Discriminant3} 
along the black line in Fig.~\ref{F:Distinct4RealRoots} 
is displayed to highlight the dynamics with one unstable fixed point at $1/2$ and two other stable fixed points that are equidistant on either side of $1/2$.  
 
We are interested in varying the externally applied stress $\sigma$ for a given material characterized by fixed rheological parameters $\alpha$ and $\beta$.  
This amounts to varying $\tilde{\sigma}$ for a fixed $\beta$ since the fixed $\alpha$ is absorbed into $\tilde{\sigma}=\sigma-\alpha$.    
The asymptotic dynamics when we apply a constant external stress for a long period of time are given by the fixed points $\overline{\mathbf{a}}^*$ in Fig.~\ref{F:Hyst}.
We study such stress-dependent behavior from the Gibbs sampler and compare it with ODE approximation in the next Section.
Note that the ODE model for $\beta \neq 0$ is only in qualitative agreement with $\overline{\mathbf{a}}(t)$, the expected volume fraction of the unyielded material at time $t$.  
This is because we are ignoring the dependent statistic $\overline{\mathbf{b}}(t)$, the expected fraction of bonds or pairs of neighboring unyielded material at time $t$.  
Despite this simplification, as we will see in Sec.~\ref{S:Results}, there is qualitative agreement between the ODE and the Gibbs simulations.  
Furthermore, an admittedly {\it ad hoc} correction of the ODE through a translation of the vector field by $(\alpha_0,\beta_0)$ even improves the quantitative approximation.  
We postpone a formal quantitative approximation of the ODE using perturbation theoretic methods to the future and focus here on obtaining insights from the Gibbs sampler that is in qualitative agreement with the ODE approximation. 

\section{Results}\label{S:Results}

In this Section we mainly obtain various insights about the macroscopic behaviour of our model based on Monte Carlo simulations from Algorithms~1 and 2 in \cite{Sainudiin15}, and make some comparisons with the approximating nonlinear ODE model of Sec.~\ref{S:ApproxModel}.

\subsection{Comparison between Microscopic model and ODE approximation under varying stress}

The energy of $X(t)$, the random site configuration at time $t$, depends on two of its highly correlated statistics: $\overline{A}(t)$, the random fraction of gelled sites at time $t$, and $\overline{B}(t)$, the random fraction of connected sites at time $t$.  
One of our primary interests is to study $\overline{A}(t)$ and $\overline{B}(t)$ as $X(t)$ is under the influence of time-varying externally applied stress $\sigma(t)$.  

\begin{figure}[htbp]
\begin{center}
\centering
\makebox{
{\includegraphics[width=0.5\textwidth]{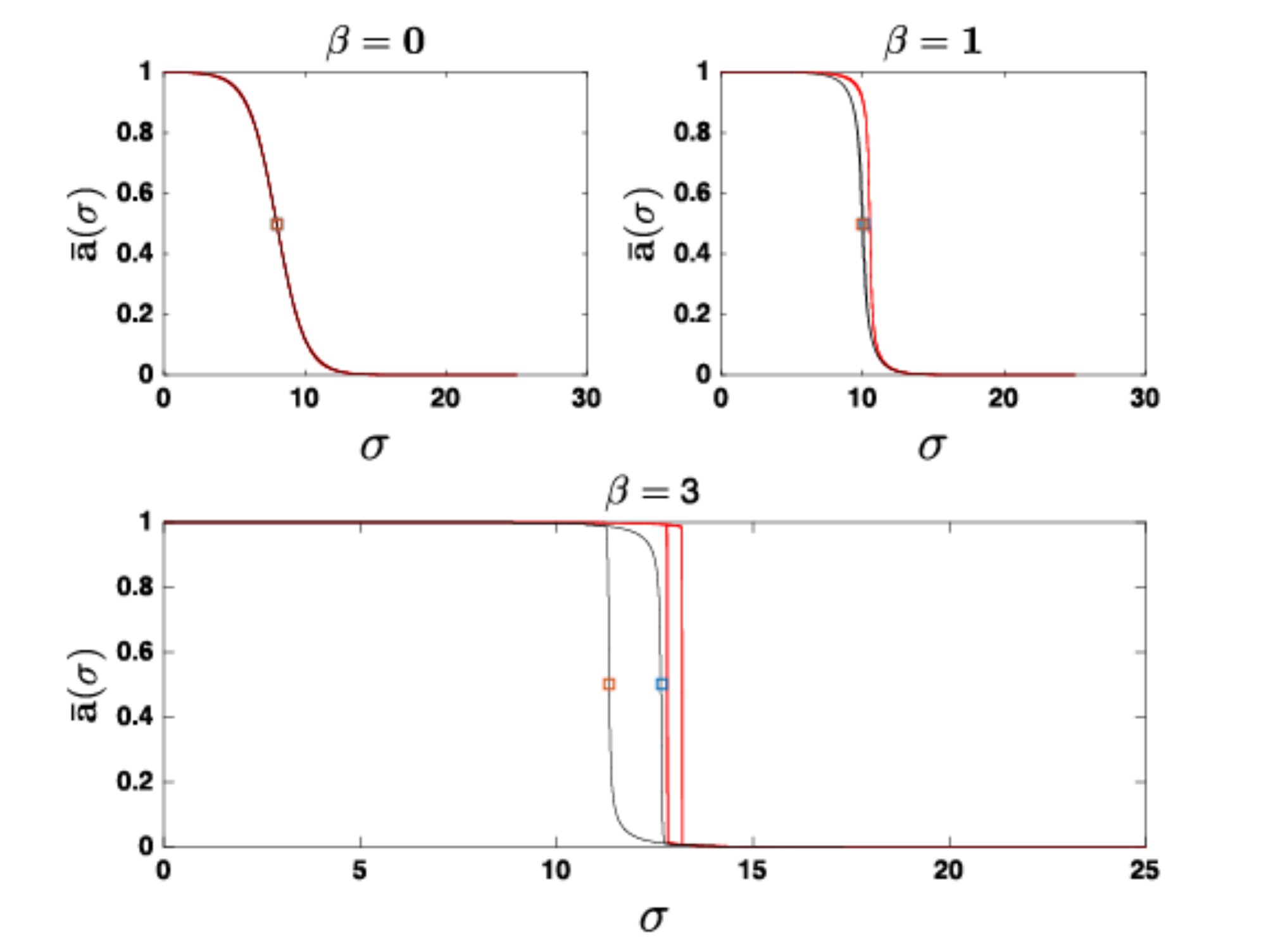}}
}
\end{center}
\caption{Gibbs field and ODE approximation simulations  with $\alpha=8$ and $\beta\in\{0,1,3\}$.  
The stress was increased from $0$ to $25$ in units of $0.01$ and decreased back to $0$ with a holding time of $t_0=1000$ (nearly asymptotic state for each distinct stress) as the site configuration varied from $\mathbf{1}$ to $\mathbf{0}$ and then back to $\mathbf{1}$. The curves with the symbol ($\square$) are the ODE simulations.
\label{F:FitFieldBeta0_1_2_4}}
\end{figure}

Using Monte Carlo simulations from Algorithm~2 in \cite{Sainudiin15} of the time-inhomogeneous Markov chain $\{X(m)\}_{m=0}^{M\underline{h}}$, under a time-dependent stress $\sigma$ ramp, we can obtain multiple independent trajectories of $\overline{A}(\sigma)$, the fraction of gelled sites as a function of the external stress $\sigma$. This is to emulate conditions of an unsteady forcing during macroscopic rhelogical measurements. In the following, $h$ is the average hits per site in the Gibbs sampler algorithm and we define it also as the characteristc forcing time $t_0$ for the stress ramp in our ODE simulations. We set $h=1000$ in order to reach steady state for each value of $\sigma$.
In Figure~\ref{F:FitFieldBeta0_1_2_4}, the trajectories are shown as thin lines and the curves for the ODE approximation have the $\square$ symbol on them.
Note the reversibility of the response of the material when $\beta\in\{0,1\}$ (top row of Figure \ref{F:FitFieldBeta0_1_2_4}) upon increasing/decreasing applied stresses. The microscopic model and the ODE approximation quantitatively agree quite well when $\beta<\beta_c$ ($\beta_c \approx 1.3$), the threshold for three fixed points in $[0,1]$ for the ODE model. As we increase $\beta$ beyond the aforementioned threshold $\beta_c$ we see that irreversible behaviour in the material appears and the comparison between the two models (discrete and continuous) is only qualitative in nature.  This is due to the fact that our ODE approximation only models $\overline{\mathbf{a}}$, instead of modelling the dependent pair $(\overline{\mathbf{a}},\overline{\mathbf{b}})$ that is sufficient for the energy, see Sec. \ref{subsec:energyconfiguration}. This effect can also bee seen if we compare the right panel of Fig.~\ref{F:Distinct4RealRoots} with Fig. 8(c) in \cite{Sainudiin15}. Clearly the light region of Fig. 8c (\cite{Sainudiin15}) corresponds to the yellow region where hysteresis is always present. The main discrepancy is the value of $\beta_c$. In our ODE approximation, the calculated value is $\beta_c\approx 1.3$, on the other hand, from our Gibbs sampler simulations $\beta_{c}^{GS}\approx1.5$. As mentioned above this difference is due to the fact that in our approximation we disregards all bond interactions between neighbours. 

As a qualitative remark one can note that even in the presence of strong interactions $\beta >\beta_c$, both models predict an increase of the steepness of the solid fluid transition (defined as the slope of the dependence $\bar a(t)$ on $\sigma$ around the point where $\bar a \approx 1/2$).

\subsection{Comparison between model by Putz and Burghelea \cite{solidfluid} and ODE approximation} 

In this section we will consider the model developed by Putz and coworkers in \cite{solidfluid,miguelstab}. This model is phenomenological in the sense that, unlike the Gibbs field model presented in Sec. \ref{S:Model} it is not derived from first principles. In this type of modelling one mimics the behaviour of the microstructure through the definition of a macroscopic structural variable with range in $[0,1]$, where $0$ means completely unstructured or fluid and 1 means completely structured or solid. The structural variable $a_p$ satisfies a kinematic equation and usually depends explicitly on the stress and/or rate of strain. In the case of \cite{solidfluid} we have:

\begin{eqnarray}
\frac{d}{dt}a_p(t)&=&k_r\left[1-\tanh\left(\frac{\sigma-\sigma_y}{w}\right)\right](1-a_p(t))\nonumber \\
&-&k_d\left[1+\tanh\left(\frac{\sigma-\sigma_y}{w}\right)\right]a_p(t). \label{PMM}
\end{eqnarray}

where $k_r$ is the rate of recombination of micro-structural units, $k_d$ is  the rate of destruction of the solid phase, $\sigma_y$ is the yield stress and $w$ is a constant that controls how steep the change in the microstructure from solid to fluid and fluid to solid is. 

\begin{figure}[htbp]
\begin{center}
\centering
\makebox{
{\includegraphics[width=0.5\textwidth]{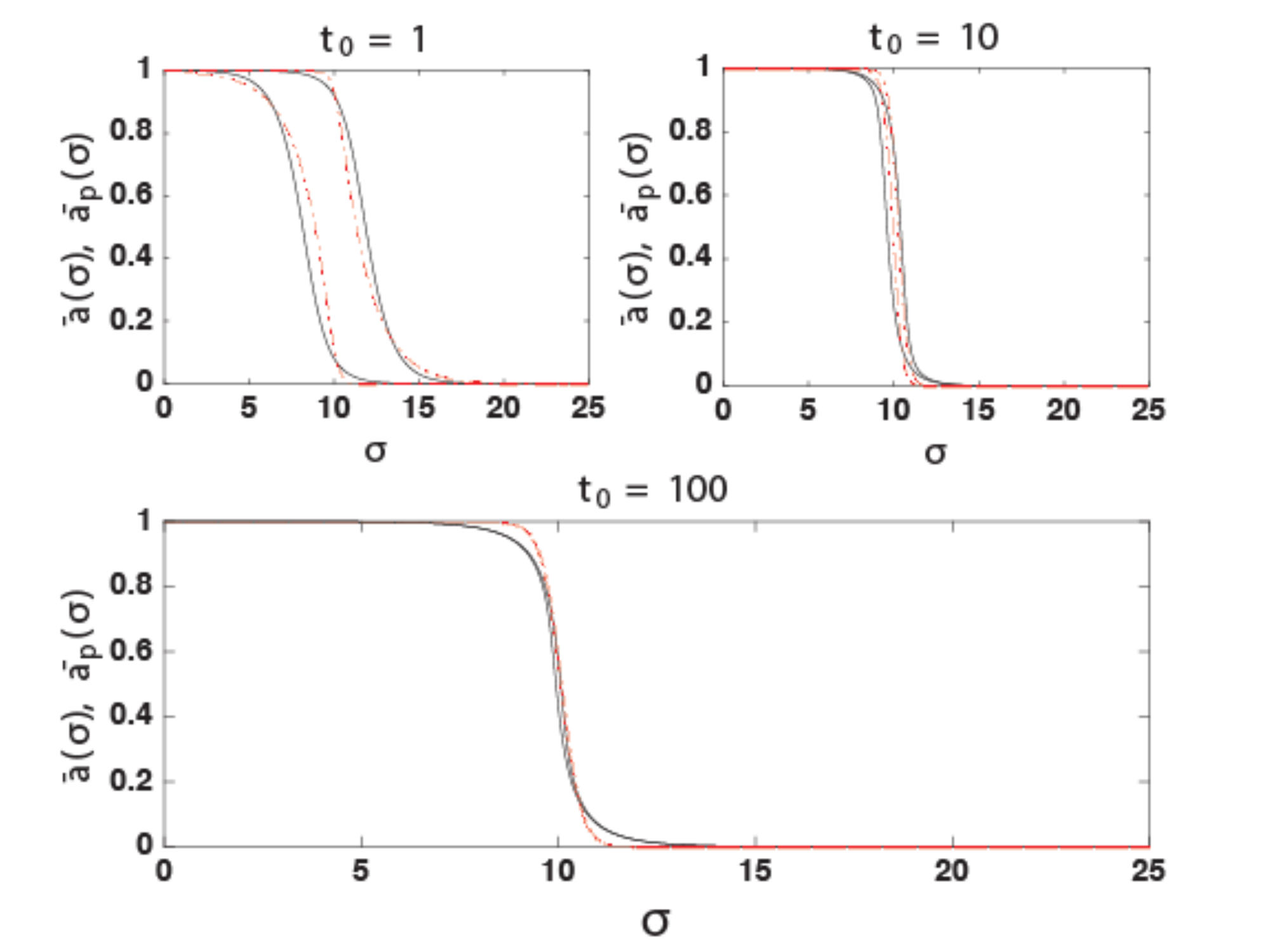}}
}
\end{center}
\caption{Comparison between ODE approxiamtion and model by Putz and Burghelea \cite{solidfluid} for different holding times $t_0$. ODE model with $\alpha=8$ and $\beta=1$, PB model with $k_d=k_r=0.3$, $w=0.5$ and $\sigma_y=10$. Full lines are the ODE approximation and broken lines the PB model.  
\label{F:ODEvsPMM}}
\end{figure}

In Figure \ref{F:ODEvsPMM} we present the simulations of equations (\ref{E:ODEAbarBetaGEQ0}) and (\ref{PMM}) for three characteristic forcing times $t_0$. As expected we have very good agreement between the models. This could be considered as a qualitative "\emph{proof}'' that the phenomenological models can actually approximate the behaviour of the microscopic models derived from first principles.

\subsection{Determination of the yield point in the limit of a steady state forcing }

A reliable estimation of the yield point is important to many practical applications involving yield stress materials.This is typically done by fitting \textit{steady state} rheological measurements with models with various degrees of complexity ranging from the mathematically simple and classical Herschel-Bulkley correlation up to structural models. Thus, it appears natural to attempt in the following to obtain an estimate of the yield point for the case of a steady state forcing from the nonlinear dynamical system model presented herein.

To get an approximation for the yield point $\sigma_y$ during a steady state forcing process we will make the assumption (well supported by the results presented in Figs. \ref{F:FitFieldBeta0_1_2_4}, \ref{F:ODEvsPMM}) that, corresponding to the yield point, the absolute value of the slope of the dependence  $\bar{a}^*(\sigma)$ passes through a maximum:

\begin{equation}\label{eq:yieldingcondition}
\left \vert \frac{d \bar{a}^*}{d \sigma}  \right \vert \overset{\sigma \approx \sigma_y}{\longmapsto} Max
\end{equation}

For simplicity, let us focus first on the non-interacting case, $\beta = 0$. 
From Eq.~\ref{E:FixedPointAbarBeta0} on can readily show that the condition \ref{eq:yieldingcondition} reduces to $\sigma_y = \alpha$. Thus, in the non-interactive case, the yield point obtained during a steady state stressing practically coincides with the site specific threshold $\alpha$ of the Gibbs field model.

\begin{figure}[]
\begin{center}
\centering
\makebox{
{\includegraphics[width=0.45\textwidth]{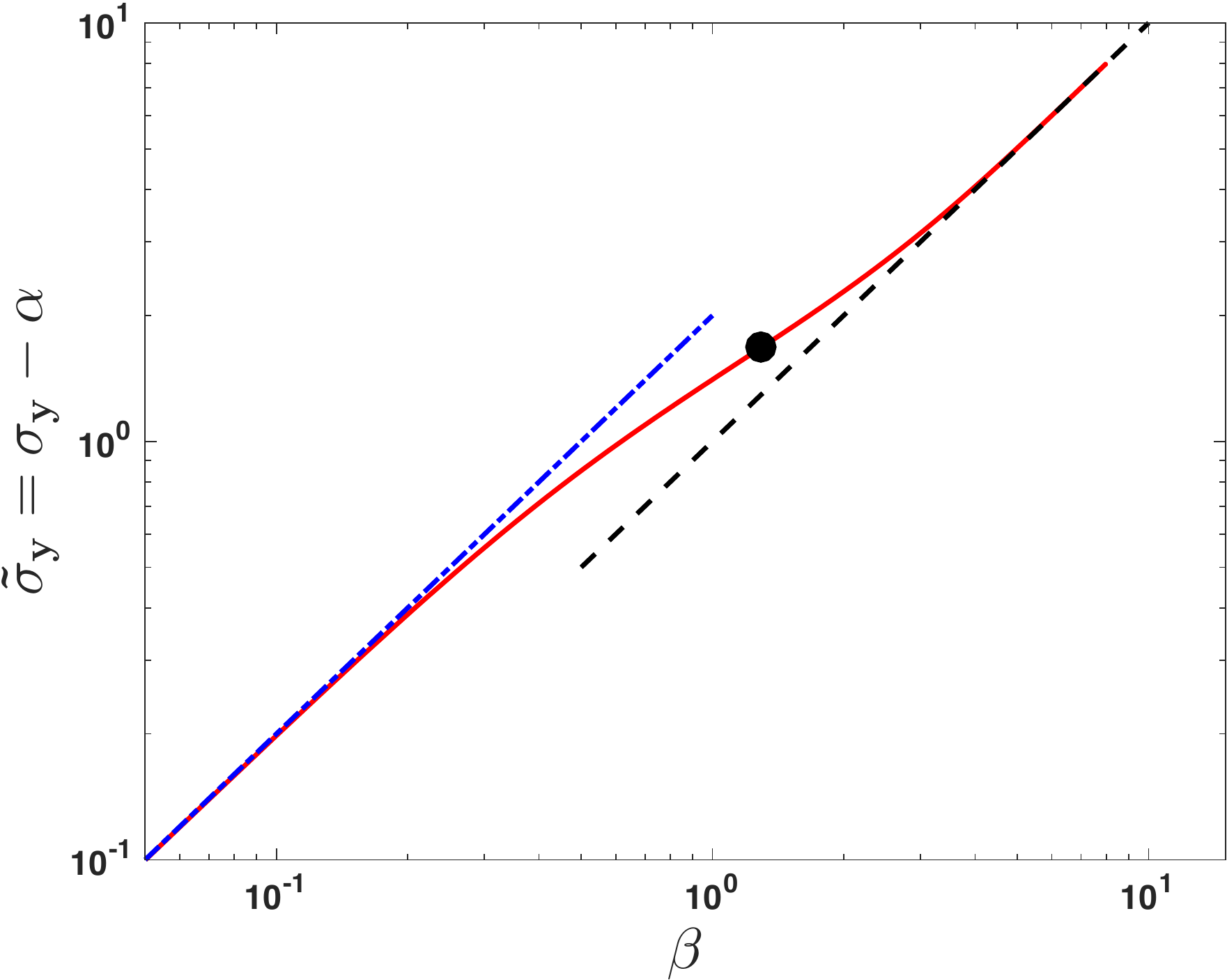}}
}
\end{center}
\caption{Dependence of the approximate yield stress shifted stress $\tilde \sigma_y = \sigma _y$ on the interaction parameter $\beta$. The dashed line is $\tilde \sigma_y  = \beta$ and the dash-dotted line is $\tilde \sigma_y  = 2 \beta$.  The circles marks the critical point corresponding to $\beta_c \approx 1.3$.
\label{F:approxyieldstress}}
\end{figure}

%

We now consider the interactive case $\beta \neq 0$. To a leading order in $\bar a^*$ and assuming that around the yield point $\bar a^* \approx 1/2$ it can be shown using Eq.~\ref{E:ODEAbarBetaGEQ0}:

\begin{equation}\label{eq:yieldstressinteracting}
\left. \frac{d \bar{a}^*}{d \tilde \sigma} \right \vert_{\sigma \approx \sigma_y}  \approx  e^{\tilde \sigma} \left[ \frac{1}{\left( 1+ e^{\tilde \sigma} \right)^2}-2 \frac{e^{-\beta}}{\left( 1+ e^{\tilde\sigma} e^{-\beta} \right)^2} \right]
\end{equation}

The implicit dependence of the approximate yield stress $\tilde \sigma_y$ on the interaction parameter $\beta$ may be obtained by solving numerically $\left \vert \frac{d \bar{a}^*)}{d \tilde \sigma}\right\vert = 0$. The result is presented in Fig. \ref{F:approxyieldstress}. For interactions weaker than the critical threshold $\beta_c$, the apparent yield stress scales as $\tilde \sigma_y = \sigma -\alpha = \beta$ (the dash-dotted line in Fig. \ref{F:approxyieldstress}). Beyond this threshold, the scaling becomes steeper, $\tilde \sigma_y = \sigma -\alpha = 2 \beta$ (the dashed line in Fig. \ref{F:approxyieldstress}). To conclude this part, the yield stress assessed via steady state controlled stress ramps is  (according to our model) expected to depend linearly on both the site specific threshold $\alpha$ which may be intuitively understood as a measure of the strength of the microscopic constituents of the fluid and the strength $\beta$ of their interaction and the slope of this behaviour switches when the strength of the interaction passes through the threshold $\beta = \beta_c$. 

\begin{figure*}
\centering
\subfigure[]{
     \includegraphics [width=0.5\textwidth] {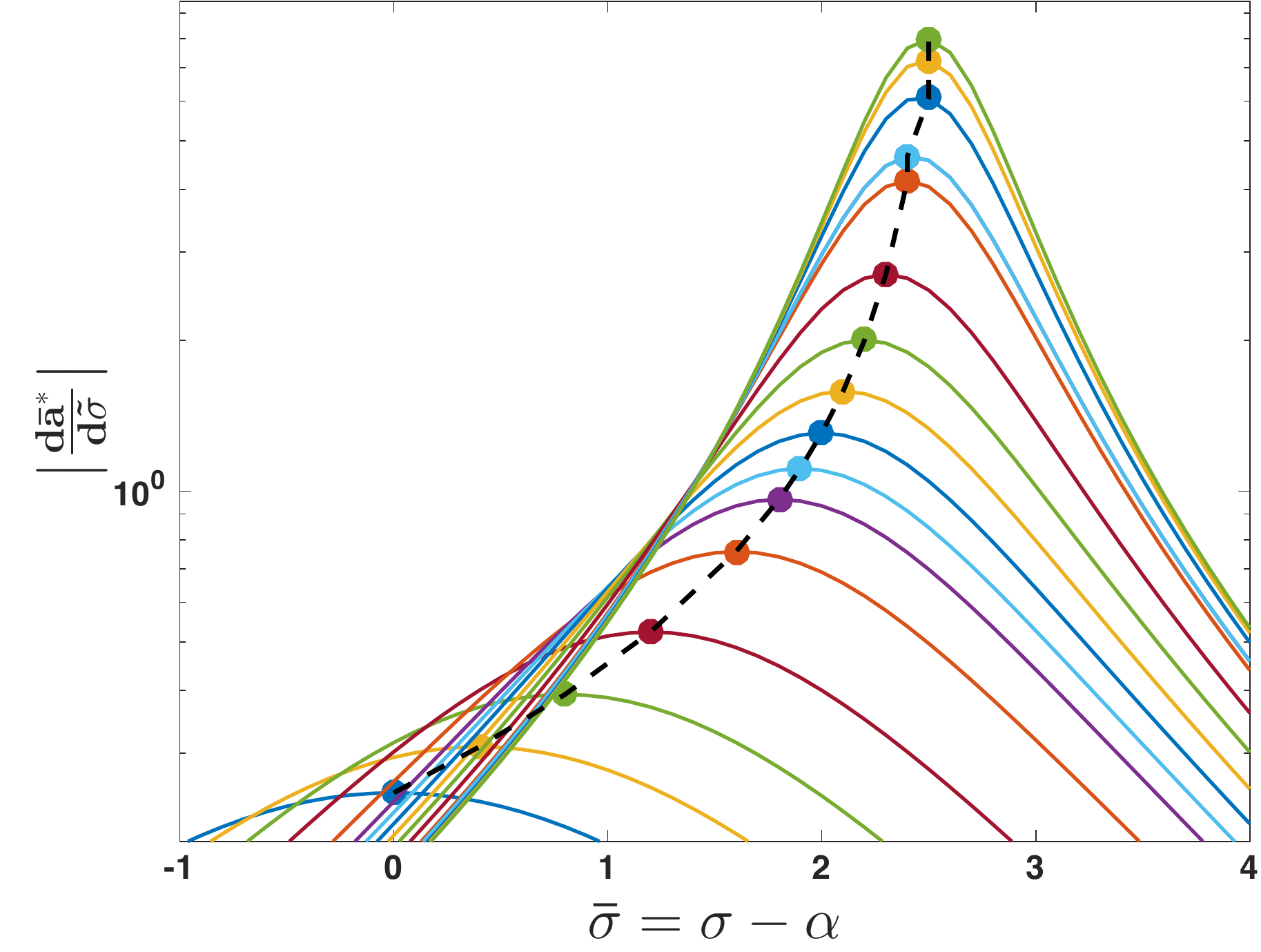}
     \label{Fig:slopes_vs_stresss}
}
\subfigure[]{
                 \includegraphics [width=0.5\textwidth] {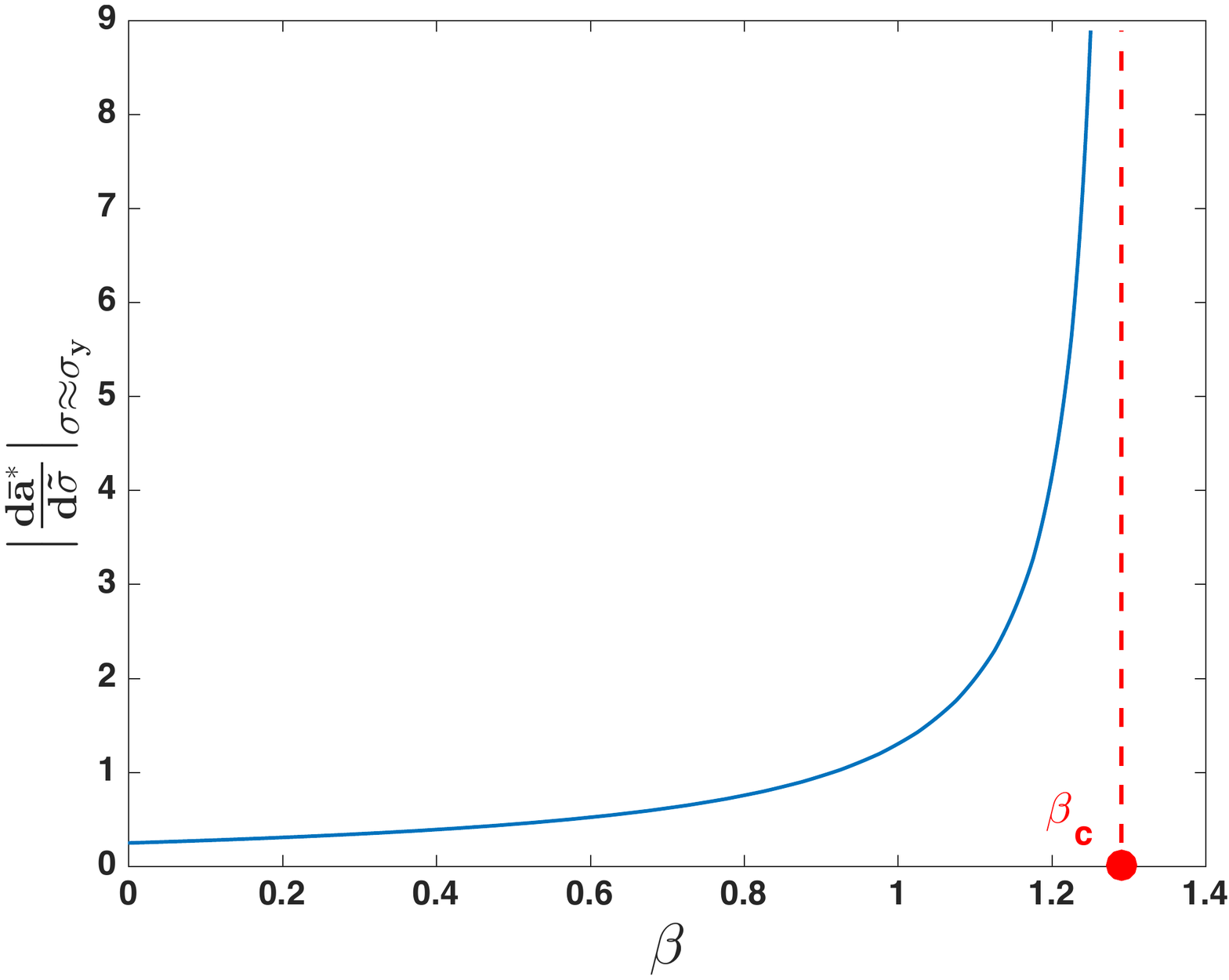}
    \label{Fig:Maxslope}}
 \caption[slopeanalysis]{\subref{Fig:slopes_vs_stresss} Dependance of the slope of the dependence of $\bar a ^*$ on the applied stress on the yield stress for various values of $\beta$ ranging from $0$ to $2$ ($\beta$ increases from bottom to top). \subref{Fig:Maxslope} Dependance of maximum value of the slope  $\left. \frac{d \bar{a}^*}{d \tilde \sigma} \right \vert_{\sigma \approx \sigma_y}$ given by Eqn.~\ref{eq:yieldingcondition} calculated around the yield point on the interaction parameter $\beta$.}
\label{fig:slopedependence}
\end{figure*}

In Fig.~\ref{fig:slopedependence} we investigate the dependence of right hand side of eqn.~\ref{eq:yieldstressinteracting} with respect to $\tilde{\sigma}$ (left panel) and with respect to $\beta$ (the right panel). 

 Regardless the value of the interaction parameter the stress dependence of the slope passes through a local maximum marked by a full symbol in Fig. \ref{Fig:slopes_vs_stresss}. As previously explained, this may be considered as an indicator of the yield point. While $\beta$ increases, the location of this maximum shifts towards larger stress values as already illustrated in Fig. \ref{F:approxyieldstress}. The value of this maximum slope increases monotonically with $beta$, the dashed line in Fig.\ref{Fig:slopes_vs_stresss}.  As we approach $\beta_c$ the slope diverges Fig. \ref{Fig:Maxslope}. This is consistent with the fact that our steady solution becomes discontinuous as a function of $\tilde{\sigma}$. Recall that we have a pitchfork bifurcation with stable fixed points $\{0,1\}$, hence the value of $\sigma_y$ is not unique and depends on the initial condition.    

\subsection{Effect of the characteristic forcing time $t_0$ on the micro-structural hysteresis}

In many practical situations the forcing time necessary to reach steady state in a given material cannot be reached. Perhaps one of the simplest such situation is that of a millimetre sized  spherical object falling slowly (speeds of order of millimetres per second) in a Carbopol gel as investigated in Ref. \cite{sedimentation}. The time scale associated to such motion is simply the ratio of the object's size to its speed and, in the case of the experiments reported in\cite{sedimentation} was of the order of  a second or shorter. To understand such hydrodynamic problems steady state rheological measurements do not suffice.
In Ref. \cite{solidfluid} the rheological response of Carbopol gel was characterised during unsteady controlled stress linear ramps for various values of the holding time $t_0$ per stress values. A rheological hysteresis (which could, at least partially and qualitatively, explain the emergence of a fore-aft symmetry breaking of the flow pattern measured for the falling sphere experiment) was systematically found and its magnitude scaled as $t_0^{-\xi}$ with $\xi \approx 1$, see Fig. 11 in \cite{solidfluid}.

The purpose of this section is to study the dependance of the area of the hysteresis of the micro-structural states observed upon increasing/decreasing forcing as a function of characteristic forcing time $t_0$, compare the results with the predictions of the Gibbs field model \cite{Sainudiin15} and with the predictions of the predictions of the structural model by Putz and Burghelea \cite{miguelstab} as well as with the experiments \cite{solidfluid,divoux4}. 
 
\begin{figure}
\begin{center}
\centering
\makebox{
{\includegraphics[width=0.5\textwidth, angle=90]{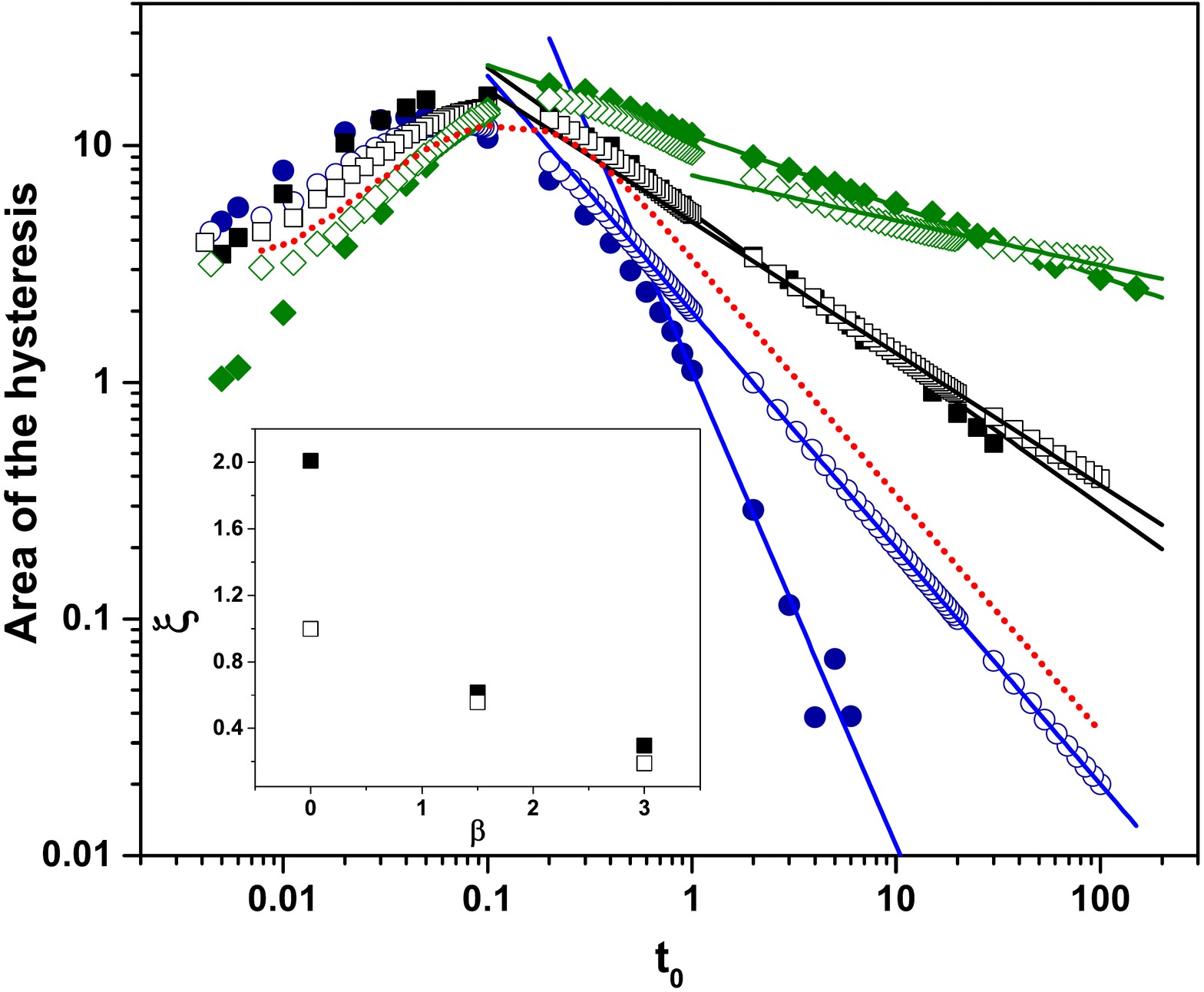}}
}
\end{center}
\caption{Comparison between Gibbs sampler simulations (GS) and ODE approximation for the change in area of the hysteresis with respect to holding time (with $\alpha=8$). The symbols refer to the value of interaction parameter $\beta$ with full symbols being the GS simulations and open symbols the ODE approximation. $\beta=0$ ($\textcolor{blue}{\bullet, \circ}$). $\beta=1.5$: ($\blacksquare$, $\square$). $\beta=3$: ($\textcolor{red}{\diamond, \blacklozenge}$). The full lines are power law fitting functions $t_0^{-\xi}$ and the exponents are presented in the insert. The dotted line is the prediction of the model by Putz and Burghelea, \cite{solidfluid}.
   \label{Fig:Areahyst}}
\end{figure}

 For this purpose, we have first run simulations using the nonlinear dynamical system described in Sec. \ref{S:ApproxModel} corresponding to several linear increasing/decreasing stress ramps for various values of the holding time per stress value $t_0$ and several values of the interaction parameter $\beta$. For each case we have calculated the area of the micro-structural hysteresis of the dependence $\bar a(\sigma)$. The results obtained of the hysteresis area on the characteristic time $t_0$ obtained from the nonlinear dynamical system approach are represented in Fig. \ref {Fig:Areahyst} as open symbols. For comparison, we calculated the same dependence by running the Gibbs field model for the same values of the interacting paramater and a number of hits per site $h$ that matches $t_0$. The results are represented in Fig. \ref {Fig:Areahyst} as full symbols. The results obtained by the two approaches are in a good qualitative agreement: in both cases the magnitude of the microstructural hysteresis depends in a non-monotonic fashion on the degree of steadiness of the external forcing. In the steady state limit of large $t_0$ ($h$), a power law scaling in the form $t_0^{-\xi}$ is observed. This finding is consistent with the experimental observations of a rheological hysteresis for a Carbopol gel subjected to an increasing/decreasing linear stress ramp in a roughened plate-plate geometry \cite{solidfluid}.
 At a quantitative level, the agreement between the results obtained via the two approaches is only partial. The biggest differences in the scaling behaviour (see the inset in  \ref {Fig:Areahyst}) are observed for the non-interactive case, $\beta = 0$ (the circles in Fig.  \ref {Fig:Areahyst}). Eventhough that for the case $\beta=0$ the approximation is exact, one should note that the Gibbs model is discrete. We expect that as $n\rightarrow\infty$ (recall that $n^2$ is the number of sites in our lattice) the rate of decay in the hysteresis will converge to the value of the continuous model. As the interaction parameter $\beta$ is gradually increased, the quantitative agreement between the two approaches improves (the circles, the squares and the rhombs). In all cases, however, the magnitude of the micro-structural hysteresis scales as a power law with the characteristic forcing time, $t_0^{-\xi}$, see the inset in Fig.  \ref {Fig:Areahyst}. It is noteworthy that the values of the scaling exponent $\xi$ are of the same order of magnitude as the ones measured experimentally in \cite{solidfluid}.  It is equally interesting to note that both approaches  predict a decrease of the scaling exponent with the interaction parameter. This finding is fully consistent with the experimental fact that, in the case of strongly interacting systems (e.g. laponite and bentonite suspensions) the rheological measurements exhibit a hysteresis (and, consequently, their reproducibility during subsequent tests is poor) even in the asymptotic case of a steady state forcing (very large waiting times $t_0$). More recently,  a large hysteresis was systematically observed even for very large values of $t_0$ during controlled stress ramps performed with a suspension of a micro-alga with an electrically charged cellular membrane, \cite{chlorella}. 

 In the case of a unsteady forcing i.e. small $t_0$ the simulations based on both approaches reveal a local maximum that shifts slightly towards larger values of $t_0$ as the interaction parameter $\beta$ increases. This finding is consistent with the experimental results obtained by Divoux and his coworkers for three materials \cite{divoux4}: a laponite suspension, for a carbon black suspension and mayonnaise. For a Carbopol gel, however, this local maximum was not observed neither in \cite{solidfluid} (see Fig.11 therein) nor in  \cite{divoux4} (see Fig. 3 (c) therein) because, most probably, it occurs at characteristic forcing time $t_0$ too small to be probed experimentally. An additional insight into the physical nature of this non monotonic behaviour was given in  \cite{Sainudiin15} by showing that for values of $t_0$ below the maximum the lattice is only partially \textit{yielded} corresponding to the maximal value of the stress reached during the ramp.
Last, we compare the predictions of the nonlinear dynamical system model with the predictions of the micro-structural model by Putz and Burghelea (the dotted line in Fig. \ref{Fig:Areahyst}), \cite{solidfluid}. One can note a fair agreement with the result obtained for the non-interactive case $\beta = 0$.     

\section{Conclusions, outlook}\label{S:Discussion}

We have presented a nonlinear dynamical system (ODE) approach for the solid fluid transition of a yield stress material subjected to an external stress that approximates the microscopic Gibbs field formulated from first principles introduced in Ref. \cite{Sainudiin15}.

In spite of some quantitative differences mainly due to the fact that the ODE approximation does not properly account for the statistics of bonds between neighbouring microscopic constituents, both the ODE and the Gibbs field approach predict several key features of the solid-transition. 

First, the transition is generally irreversible upon increasing/decreasing forcing an a micro-structural hysteresis is systematically observed, Fig. \ref{F:FitFieldBeta0_1_2_4}. A reversible transition may be observed solely in the non-interacting case $\beta = 0$ and in the limit of a steady state forcing (the top left panel in Fig. \ref{F:FitFieldBeta0_1_2_4}). By a systematic analysis of the stability of the fixed points of the nonlinear dynamical system we could show that a genuine hysteresis will be observed even in the asymptotic limit of steady forcing if the interaction parameter exceeds the threshold $\beta_c \approx 1.3$, (bottom panel in Fig. \ref{F:FitFieldBeta0_1_2_4} ). The magnitude of the micro-structural hysteresis depends on both the level of interactions between the microscopic constituents and the degree of steadiness of the external forcing $t_0$. In the limit of slow forcing the hysteresis decays as a power law $t_0^{-\xi}$, Fig. \ref{Fig:Areahyst} and the power law exponents decreases with increasing strength of the interactions, the inset in Fig. \ref{Fig:Areahyst}. The monotonic decrease of the scaling exponent with $\beta$ implies that, when strong interactions among the microscopic constituents are present, a strong irreversibility of the micro-structural states upon increasing/decreasing stresses even in the asymptotic limit of a steady state forcing.   

Second, the abruptness of the solid fluid gradually increases with increasing interaction parameter $\beta$, Fig. \ref{F:FitFieldBeta0_1_2_4}.

Third, we remark that the Gibbs Simulations as well as the approximating ODE are in qualitative agreement with the simple phenomenological  model for the micro-structural hysteresis proposed in \ref{F:ODEvsPMM}.

Fourth and equally important from a practical perspective, our model allows one to estimate the yield point and monitor its behaviour as a function of the interaction parameter, Fig. \ref{F:approxyieldstress}. A linear dependence is found and its slope changes corresponding to the critical point $\beta = \beta_c$. When $\beta>\beta_c$ the value of the yield stress is no longer unique. This due to the fact that the process is no longer reversible. The steepness of the yielding transition diverges at the critical point $\beta = \beta_c$, Fig. \ref{fig:slopedependence}.

In closing, we believe there are several future directions worth pursuing. 
 At a theoretical level, a more quantitative comparison between these models aimed at highlighting their differences may be useful.  
Ideally, perturbation theoretic methods should be used to improve the quantitative agreement between the nonlinear ODE model and the stochastic trajectories as opposed to the ad-hoc translations of the vector field done in this study. 
A more detailed model that simultaneously represents the fraction of gelled sites and the fraction of bonds in one dependent system would provide a better quantitative and qualitative approximation of the correlated site percolation model. Another interesting extension of our model could involve allowing for solvent effects through a model akin to {\it correlated site-bond percolation} of \cite[Sec.~D.II., p.136]{Stauffer1982} but with our focus on external stress as opposed to temperature.  
In such a model we have an additional parameter that allows for a site to be occupied by a monomer with probability $\phi$ and by the solvent with probability $1-\phi$.  

From the more practical standpoint of the rheologist, it would be interesting to couple the nonlinear dynamical system approach to an appropriate elasto-viscoplastic constitutive relation and attempt either fitting experimental data or modelling industrially relevant non rheometric flow problems.
  
\section*{Acknowledgements}
This collaborative work commenced during TB's visit to University of Canterbury (UC) in 2012 with partial support from a research grant to MM-G from UC's College of Engineering and by the project ThIM to TB from the National French Research Agency (ANR). 
RS and MM-G visited Nantes in 2013 with partial support from the project ThIM, the Laboratoire de Thermocin\'etique (LTN UMR 6607) and research travel grants from UC's College of Engineering.  
RS thanks Anusha Raazesh for diagrams of neighborhood configurations that led to the binomial approximation in Section~\ref{S:binomApprox} and Brendan Creutz for discussions on parametrized discriminants of cubic and quartic polynomials.


\begin{thebibliography}{54}
\expandafter\ifx\csname natexlab\endcsname\relax\def\natexlab#1{#1}\fi
\providecommand{\bibinfo}[2]{#2}
\ifx\xfnm\relax \def\xfnm[#1]{\unskip,\space#1}\fi
\bibitem[{R.~Sainudiin and Burghelea(2015)}]{Sainudiin15}
\bibinfo{author}{M.~M.-G. R.~Sainudiin}, \bibinfo{author}{T.~Burghelea},
\newblock \bibinfo{title}{A microscopic {G}ibbs field model for the macroscopic
  yielding behaviour of a viscoplastic fluid},
\newblock \bibinfo{journal}{Soft Matter} \bibinfo{volume}{11 (27)}
  (\bibinfo{year}{2015}) \bibinfo{pages}{5531--5545}.
\bibitem[{Putz and Burghelea(2009)}]{solidfluid}
\bibinfo{author}{A.~M.~V. Putz}, \bibinfo{author}{T.~I. Burghelea},
\newblock \bibinfo{title}{The solid-fluid transition in a yield stress shear
  thinning physical gel},
\newblock \bibinfo{journal}{Rheol Acta} \bibinfo{volume}{48}
  (\bibinfo{year}{2009}) \bibinfo{pages}{673--689}.
\bibitem[{Han et~al.(1997)Han, Sung, and Kim}]{jeong}
\bibinfo{author}{B.~Y. Han}, \bibinfo{author}{L.~D. Sung},
  \bibinfo{author}{S.~W. Kim},
\newblock \bibinfo{title}{Biodegradable block copolymers as injectable
  drug-delivery systems},
\newblock \bibinfo{journal}{Nature} \bibinfo{volume}{388}
  (\bibinfo{year}{1997}) \bibinfo{pages}{860--862}.
\bibitem[{Qiu and Park(2001)}]{Qiu2001321}
\bibinfo{author}{Y.~Qiu}, \bibinfo{author}{K.~Park},
\newblock \bibinfo{title}{Environment-sensitive hydrogels for drug delivery},
\newblock \bibinfo{journal}{Advanced Drug Delivery Reviews}
  \bibinfo{volume}{53} (\bibinfo{year}{2001}) \bibinfo{pages}{321 -- 339}.
  \bibinfo{note}{Triggering in Drug Delivery Systems}.
\bibitem[{Hou et~al.(2004)Hou, Bank, and Shakesheff}]{hou}
\bibinfo{author}{Q.~Hou}, \bibinfo{author}{P.~A.~D. Bank},
  \bibinfo{author}{K.~M. Shakesheff},
\newblock \bibinfo{title}{Injectable scaffolds for tissue regeneration},
\newblock \bibinfo{journal}{J. Matter. Chem.} \bibinfo{volume}{14}
  (\bibinfo{year}{2004}) \bibinfo{pages}{1915}.
\bibitem[{Beck et~al.(2007)Beck, Madsen, Britt, Vernon, and Nguyen}]{beck}
\bibinfo{author}{J.~Beck}, \bibinfo{author}{B.~Madsen},
  \bibinfo{author}{D.~Britt}, \bibinfo{author}{B.~Vernon},
  \bibinfo{author}{K.~T. Nguyen},
\newblock \bibinfo{title}{Islet encapsulation: Strategies to enhance islet cell
  functions},
\newblock \bibinfo{journal}{Tissue Engineering} \bibinfo{volume}{13}
  (\bibinfo{year}{2007}) \bibinfo{pages}{589--599}.
\bibitem[{Nguyen and Boger(1992)}]{bogerreview}
\bibinfo{author}{Q.~D. Nguyen}, \bibinfo{author}{D.~V. Boger},
\newblock \bibinfo{title}{Measuring the flow properties of yield stress
  fluids},
\newblock \bibinfo{journal}{Annual Review of Fluid Mechanics}
  \bibinfo{volume}{24} (\bibinfo{year}{1992}) \bibinfo{pages}{47--88}.
\bibitem[{Coussot(2014)}]{Coussot201431}
\bibinfo{author}{P.~Coussot},
\newblock \bibinfo{title}{Yield stress fluid flows: A review of experimental
  data},
\newblock \bibinfo{journal}{Journal of Non-Newtonian Fluid Mechanics}
  \bibinfo{volume}{211} (\bibinfo{year}{2014}) \bibinfo{pages}{31 -- 49}.
\bibitem[{Balmforth et~al.(2014)Balmforth, Frigaard, and Ovarlez}]{ianreview}
\bibinfo{author}{N.~J. Balmforth}, \bibinfo{author}{I.~A. Frigaard},
  \bibinfo{author}{G.~Ovarlez},
\newblock \bibinfo{title}{Yielding to stress: Recent developments in
  viscoplastic fluid mechanics},
\newblock \bibinfo{journal}{Annual Review of Fluid Mechanics}
  \bibinfo{volume}{46} (\bibinfo{year}{2014}) \bibinfo{pages}{121--146}.
\bibitem[{Bonn et~al.(2015)Bonn, Paredes, Denn, Berthier, Divoux, and
  Manneville}]{dennreview}
\bibinfo{author}{D.~Bonn}, \bibinfo{author}{J.~Paredes}, \bibinfo{author}{M.~M.
  Denn}, \bibinfo{author}{L.~Berthier}, \bibinfo{author}{T.~Divoux},
  \bibinfo{author}{S.~Manneville},
\newblock \bibinfo{title}{Yield stress materials in soft condensed matter},
\newblock \bibinfo{journal}{arXiv:1502.05281 [cond-mat.soft]}
  (\bibinfo{year}{2015}).
\bibitem[{Barnes(1999)}]{barnes1}
\bibinfo{author}{H.~A. Barnes},
\newblock \bibinfo{title}{The yield stress--a review or
  `{$\pi\alpha\nu\tau\alpha \ \rho\epsilon\iota$}'--everything flows?},
\newblock \bibinfo{journal}{Journal of Non-Newtonian Fluid Mechanics}
  \bibinfo{volume}{81} (\bibinfo{year}{1999}) \bibinfo{pages}{133--178}.
\bibitem[{Barnes and Walters(1985)}]{barnes2}
\bibinfo{author}{H.~A. Barnes}, \bibinfo{author}{K.~Walters},
\newblock \bibinfo{title}{The yield stress myth?},
\newblock \bibinfo{journal}{Rheol. Acta} \bibinfo{volume}{24}
  (\bibinfo{year}{1985}) \bibinfo{pages}{323--326}.
\bibitem[{Moller et~al.(2009)Moller, Fall, and Bonn}]{moller1}
\bibinfo{author}{P.~C.~F. Moller}, \bibinfo{author}{A.~Fall},
  \bibinfo{author}{D.~Bonn},
\newblock \bibinfo{title}{Origin of apparent viscosity in yield stress fluids
  below yielding},
\newblock \bibinfo{journal}{EPL (Europhysics Letters)} \bibinfo{volume}{87}
  (\bibinfo{year}{2009}) \bibinfo{pages}{38004}.
\bibitem[{Bonn and Denn(2009)}]{dennyieldstress}
\bibinfo{author}{D.~Bonn}, \bibinfo{author}{M.~M. Denn},
\newblock \bibinfo{title}{Yield stress fluids slowly yield to analysis},
\newblock \bibinfo{journal}{Science} \bibinfo{volume}{324}
  (\bibinfo{year}{2009}) \bibinfo{pages}{1401--1402}.
\bibitem[{Denn and Bonn(2011)}]{dennrheoacta}
\bibinfo{author}{M.~Denn}, \bibinfo{author}{D.~Bonn},
\newblock \bibinfo{title}{Issues in the flow of yield-stress liquids},
\newblock \bibinfo{journal}{Rheologica Acta} \bibinfo{volume}{50}
  (\bibinfo{year}{2011}) \bibinfo{pages}{307--315}.
  \bibinfo{note}{10.1007/s00397-010-0504-3}.
\bibitem[{Herschel and Bulkley(1926{\natexlab{a}})}]{originalhb}
\bibinfo{author}{W.~H. Herschel}, \bibinfo{author}{R.~Bulkley},
\newblock \bibinfo{title}{Konsistenzmessungen von gummi-benzoll\"{o}sungen},
\newblock \bibinfo{journal}{Kolloid-Zeitschrift} \bibinfo{volume}{39}
  (\bibinfo{year}{1926}{\natexlab{a}}) \bibinfo{pages}{291--300}.
\bibitem[{Herschel and Bulkley(1926{\natexlab{b}})}]{H-B}
\bibinfo{author}{W.~Herschel}, \bibinfo{author}{T.~Bulkley},
\newblock \bibinfo{title}{Measurement of consistency as applied to
  rubber–benzene solutions},
\newblock \bibinfo{journal}{Am. Soc. Test Proc.} \bibinfo{volume}{26(2)}
  (\bibinfo{year}{1926}{\natexlab{b}}) \bibinfo{pages}{621–633}.
\bibitem[{Papanastasiou(1987)}]{Papanastasiou1987}
\bibinfo{author}{T.~C. Papanastasiou},
\newblock \bibinfo{title}{Flows of materials with yield},
\newblock \bibinfo{journal}{Journal of Rheology (1978-present)}
  \bibinfo{volume}{31} (\bibinfo{year}{1987}) \bibinfo{pages}{385--404}.
\bibitem[{Tropea et~al.(2007)Tropea, Yarin, and Foss}]{handbook}
\bibinfo{author}{C.~Tropea}, \bibinfo{author}{A.~L. Yarin},
  \bibinfo{author}{J.~S. Foss}, \bibinfo{title}{Handbook of experimental fluid
  dynamics}, \bibinfo{publisher}{Springer-Verlag, Berlin Heidelberg},
  \bibinfo{year}{2007}.
\bibitem[{M\"{o}ller et~al.(2006)M\"{o}ller, Mewis, and Bonn}]{moller}
\bibinfo{author}{C.~F. M\"{o}ller, Peder}, \bibinfo{author}{J.~Mewis},
  \bibinfo{author}{D.~Bonn},
\newblock \bibinfo{title}{Yield stress and thixotropy: on the difficulty of
  measuring yield stress in practice},
\newblock \bibinfo{journal}{Soft Matter} \bibinfo{volume}{2}
  (\bibinfo{year}{2006}) \bibinfo{pages}{274--283}.
\bibitem[{Moller et~al.(2009)Moller, Fall, Chikkadi, Derks, and
  Bonn}]{Moller5139}
\bibinfo{author}{P.~Moller}, \bibinfo{author}{A.~Fall},
  \bibinfo{author}{V.~Chikkadi}, \bibinfo{author}{D.~Derks},
  \bibinfo{author}{D.~Bonn},
\newblock \bibinfo{title}{An attempt to categorize yield stress fluid
  behaviour},
\newblock \bibinfo{journal}{Philosophical Transactions of the Royal Society of
  London A: Mathematical, Physical and Engineering Sciences}
  \bibinfo{volume}{367} (\bibinfo{year}{2009}) \bibinfo{pages}{5139--5155}.
\bibitem[{Moyers-Gonzalez et~al.(2011)Moyers-Gonzalez, Burghelea, and
  Mak}]{miguelstab}
\bibinfo{author}{M.~Moyers-Gonzalez}, \bibinfo{author}{T.~I. Burghelea},
  \bibinfo{author}{J.~Mak},
\newblock \bibinfo{title}{Linear stability analysis for plane-poiseuille flow
  of an elastoviscoplastic fluid with internal microstructure for large
  reynolds numbers},
\newblock \bibinfo{journal}{Journal of Non-Newtonian Fluid Mechanics}
  \bibinfo{volume}{166} (\bibinfo{year}{2011}) \bibinfo{pages}{515 -- 531}.
\bibitem[{Weber et~al.(2012)Weber, Moyers-Gonzalez, and Burghelea}]{thermo}
\bibinfo{author}{E.~Weber}, \bibinfo{author}{M.~Moyers-Gonzalez},
  \bibinfo{author}{T.~I. Burghelea},
\newblock \bibinfo{title}{Thermorheological properties of a carbopol gel under
  shear},
\newblock \bibinfo{journal}{Journal of Non-Newtonian Fluid Mechanics}
  \bibinfo{volume}{183-184} (\bibinfo{year}{2012}) \bibinfo{pages}{14 -- 24}.
\bibitem[{Divoux et~al.(2011)Divoux, Barentin, and Manneville}]{divoux3}
\bibinfo{author}{T.~Divoux}, \bibinfo{author}{C.~Barentin},
  \bibinfo{author}{S.~Manneville},
\newblock \bibinfo{title}{From stress-induced fluidization processes to
  herschel-bulkley behaviour in simple yield stress fluids},
\newblock \bibinfo{journal}{Soft Matter} \bibinfo{volume}{7}
  (\bibinfo{year}{2011}) \bibinfo{pages}{8409--8418}.
\bibitem[{Divoux et~al.(2013)Divoux, Grenard, and Manneville}]{divoux4}
\bibinfo{author}{T.~Divoux}, \bibinfo{author}{V.~Grenard},
  \bibinfo{author}{S.~Manneville},
\newblock \bibinfo{title}{Rheological hysteresis in soft glassy materials},
\newblock \bibinfo{journal}{Phys. Rev. Lett.} \bibinfo{volume}{110}
  (\bibinfo{year}{2013}) \bibinfo{pages}{018304}.
\bibitem[{Putz et~al.(2008)Putz, Burghelea, Frigaard, and
  Martinez}]{sedimentation}
\bibinfo{author}{A.~M.~V. Putz}, \bibinfo{author}{T.~I. Burghelea},
  \bibinfo{author}{I.~A. Frigaard}, \bibinfo{author}{D.~M. Martinez},
\newblock \bibinfo{title}{Settling of an isolated spherical particle in a yield
  stress shear thinning fluid.},
\newblock \bibinfo{journal}{Phys. Fluids}  (\bibinfo{year}{2008})
  \bibinfo{pages}{033102}.
\bibitem[{Poumaere et~al.(2014)Poumaere, Moyers-González, Castelain, and
  Burghelea}]{unsteady}
\bibinfo{author}{A.~Poumaere}, \bibinfo{author}{M.~Moyers-González},
  \bibinfo{author}{C.~Castelain}, \bibinfo{author}{T.~Burghelea},
\newblock \bibinfo{title}{Unsteady laminar flows of a carbopol® gel in the
  presence of wall slip},
\newblock \bibinfo{journal}{Journal of Non-Newtonian Fluid Mechanics}
  \bibinfo{volume}{205} (\bibinfo{year}{2014}) \bibinfo{pages}{28 -- 40}.
\bibitem[{Kebiche et~al.(2014)Kebiche, Castelain, and
  Burghelea}]{teoconvection}
\bibinfo{author}{Z.~Kebiche}, \bibinfo{author}{C.~Castelain},
  \bibinfo{author}{T.~Burghelea},
\newblock \bibinfo{title}{Experimental investigation of the
  rayleigh--b{\'e}nard convection in a yield stress fluid},
\newblock \bibinfo{journal}{Journal of Non-Newtonian Fluid Mechanics}
  \bibinfo{volume}{203} (\bibinfo{year}{2014}) \bibinfo{pages}{9 -- 23}.
\bibitem[{Dullaert and Mewis(2006)}]{dullaert}
\bibinfo{author}{K.~Dullaert}, \bibinfo{author}{J.~Mewis},
\newblock \bibinfo{title}{A structural kinetics model for thixotropy},
\newblock \bibinfo{journal}{J. Non-Newtonian Fluid Mech.}
  (\bibinfo{year}{2006}) \bibinfo{pages}{21--30}.
\bibitem[{Quemada(1998{\natexlab{a}})}]{quemada1}
\bibinfo{author}{D.~Quemada},
\newblock \bibinfo{title}{Rheological modeling of complex fluids: {I}:{T}he
  concept of effective volume fraction revisited},
\newblock \bibinfo{journal}{Eur. Phys. J. AP}
  (\bibinfo{year}{1998}{\natexlab{a}}) \bibinfo{pages}{119--127}.
\bibitem[{Quemada(1998{\natexlab{b}})}]{quemada3}
\bibinfo{author}{D.~Quemada},
\newblock \bibinfo{title}{Rheological modeling of complex fluids: {III}:
  {D}ilatant behaviour of stabilized suspensions},
\newblock \bibinfo{journal}{Eur. Phys. J. AP}
  (\bibinfo{year}{1998}{\natexlab{b}}) \bibinfo{pages}{309--320}.
\bibitem[{Quemada(1999)}]{quemada4}
\bibinfo{author}{D.~Quemada},
\newblock \bibinfo{title}{Rheological modeling of complex fluids: {IV}:
  {T}hixotropic and "thixoelastic" behaviour. {S}tart-up and stress relaxation,
  creep tests and hysteresis cycles},
\newblock \bibinfo{journal}{Eur. Phys. J. AP}  (\bibinfo{year}{1999})
  \bibinfo{pages}{191--207}.
\bibitem[{Coussot et~al.(2002{\natexlab{a}})Coussot, Nguyen, Huynh, and
  Bonn}]{avalanchecoussot}
\bibinfo{author}{P.~Coussot}, \bibinfo{author}{Q.~D. Nguyen},
  \bibinfo{author}{H.~T. Huynh}, \bibinfo{author}{D.~Bonn},
\newblock \bibinfo{title}{Avalanche behavior in yield stress fluids},
\newblock \bibinfo{journal}{Phys. Rev. Lett.} \bibinfo{volume}{88}
  (\bibinfo{year}{2002}{\natexlab{a}}) \bibinfo{pages}{175501}.
\bibitem[{Coussot et~al.(2002{\natexlab{b}})Coussot, Nguyen, Huynh, and
  Bonn}]{avalanchecoussot1}
\bibinfo{author}{P.~Coussot}, \bibinfo{author}{Q.~D. Nguyen},
  \bibinfo{author}{H.~T. Huynh}, \bibinfo{author}{D.~Bonn},
\newblock \bibinfo{title}{Viscosity bifurcation in thixotropic, yielding
  fluids},
\newblock \bibinfo{journal}{Journal of Rheology} \bibinfo{volume}{46}
  (\bibinfo{year}{2002}{\natexlab{b}}) \bibinfo{pages}{573--589}.
\bibitem[{Roussel et~al.(2004)Roussel, Le~Roy, and Coussot}]{coussotthixotropy}
\bibinfo{author}{N.~Roussel}, \bibinfo{author}{R.~Le~Roy},
  \bibinfo{author}{P.~Coussot},
\newblock \bibinfo{title}{Thixotropy modelling at local and macrsocopic
  scales},
\newblock \bibinfo{journal}{J. non-Newtonian Fluid Mech.} \bibinfo{volume}{117}
  (\bibinfo{year}{2004}) \bibinfo{pages}{85--95}.
\bibitem[{de~Souza~Mendes(2009)}]{paulo1}
\bibinfo{author}{P.~R. de~Souza~Mendes},
\newblock \bibinfo{title}{Modeling the thixotropic behavior of structured
  fluids},
\newblock \bibinfo{journal}{Journal of Non-Newtonian Fluid Mechanics}
  \bibinfo{volume}{164} (\bibinfo{year}{2009}) \bibinfo{pages}{66 -- 75}.
\bibitem[{de~Souza~Mendes(2011)}]{paulo2}
\bibinfo{author}{P.~R. de~Souza~Mendes},
\newblock \bibinfo{title}{Thixotropic elasto-viscoplastic model for structured
  fluids},
\newblock \bibinfo{journal}{Soft Matter} \bibinfo{volume}{7}
  (\bibinfo{year}{2011}) \bibinfo{pages}{2471--2483}.
\bibitem[{Dimitriou et~al.(2013)Dimitriou, Ewoldt, and
  McKinley}]{garethyielding}
\bibinfo{author}{C.~J. Dimitriou}, \bibinfo{author}{R.~H. Ewoldt},
  \bibinfo{author}{G.~H. McKinley},
\newblock \bibinfo{title}{Describing and prescribing the constitutive response
  of yield stress fluids using large amplitude oscillatory shear stress
  (laostress)},
\newblock \bibinfo{journal}{Journal of Rheology (1978-present)}
  \bibinfo{volume}{57} (\bibinfo{year}{2013}) \bibinfo{pages}{27--70}.
\bibitem[{Dimitriou and McKinley(2014)}]{garethlaos}
\bibinfo{author}{C.~J. Dimitriou}, \bibinfo{author}{G.~H. McKinley},
\newblock \bibinfo{title}{A comprehensive constitutive law for waxy crude oil:
  a thixotropic yield stress fluid},
\newblock \bibinfo{journal}{Soft Matter} \bibinfo{volume}{10}
  (\bibinfo{year}{2014}) \bibinfo{pages}{6619--6644}.
\bibitem[{Blackwell and Ewoldt(2014)}]{randymodel}
\bibinfo{author}{B.~C. Blackwell}, \bibinfo{author}{R.~H. Ewoldt},
\newblock \bibinfo{title}{A simple thixotropic-viscoelastic constitutive model
  produces unique signatures in large-amplitude oscillatory shear (laos)},
\newblock \bibinfo{journal}{Journal of Non-Newtonian Fluid Mechanics}
  \bibinfo{volume}{208?209} (\bibinfo{year}{2014}) \bibinfo{pages}{27 -- 41}.
\bibitem[{Picard et~al.(2002)Picard, Ajdari, Bocquet, and
  Lequeux}]{picardfluidity}
\bibinfo{author}{G.~Picard}, \bibinfo{author}{A.~Ajdari},
  \bibinfo{author}{L.~Bocquet}, \bibinfo{author}{F.~m.~c. Lequeux},
\newblock \bibinfo{title}{Simple model for heterogeneous flows of yield stress
  fluids},
\newblock \bibinfo{journal}{Phys. Rev. E} \bibinfo{volume}{66}
  (\bibinfo{year}{2002}) \bibinfo{pages}{051501}.
\bibitem[{Bautista et~al.(2009)Bautista, Munoz, Castillo-Tejas,
  P\'{e}rez-L\'{o}pez, Puig, and Manero}]{manero}
\bibinfo{author}{F.~Bautista}, \bibinfo{author}{M.~Munoz},
  \bibinfo{author}{J.~Castillo-Tejas}, \bibinfo{author}{J.~H.
  P\'{e}rez-L\'{o}pez}, \bibinfo{author}{J.~E. Puig},
  \bibinfo{author}{O.~Manero},
\newblock \bibinfo{title}{Critical phenomenon analysis of shear-banding flow in
  polymer-like micellar solutions. 1. theoretical approach},
\newblock \bibinfo{journal}{The Journal of Physical Chemistry B}
  \bibinfo{volume}{113} (\bibinfo{year}{2009}) \bibinfo{pages}{16101--16109}.
  \bibinfo{note}{PMID: 19924843}.
\bibitem[{Hong et~al.(2008)Hong, Zhao, Zhou, and Suo}]{Hong20081779}
\bibinfo{author}{W.~Hong}, \bibinfo{author}{X.~Zhao},
  \bibinfo{author}{J.~Zhou}, \bibinfo{author}{Z.~Suo},
\newblock \bibinfo{title}{A theory of coupled diffusion and large deformation
  in polymeric gels},
\newblock \bibinfo{journal}{Journal of the Mechanics and Physics of Solids}
  \bibinfo{volume}{56} (\bibinfo{year}{2008}) \bibinfo{pages}{1779 -- 1793}.
\bibitem[{An et~al.(2010)An, Solis, and Jiang}]{solis}
\bibinfo{author}{Y.~An}, \bibinfo{author}{F.~J. Solis},
  \bibinfo{author}{H.~Jiang},
\newblock \bibinfo{title}{A thermodynamic model of physical gels},
\newblock \bibinfo{journal}{Journal of the Mechanics and Physics of Solids}
  \bibinfo{volume}{58} (\bibinfo{year}{2010}) \bibinfo{pages}{2083 -- 2099}.
\bibitem[{Peshkov et~al.(2014)Peshkov, Grmela, and Romenski}]{Peshkov2014}
\bibinfo{author}{I.~Peshkov}, \bibinfo{author}{M.~Grmela},
  \bibinfo{author}{E.~Romenski},
\newblock \bibinfo{title}{Irreversible mechanics and thermodynamics of
  two-phase continua experiencing stress-induced solid--fluid transitions},
\newblock \bibinfo{journal}{Continuum Mechanics and Thermodynamics}
  \bibinfo{volume}{27} (\bibinfo{year}{2014}) \bibinfo{pages}{905--940}.
\bibitem[{de~Bruyn(2013)}]{debruyn}
\bibinfo{author}{J.~R. de~Bruyn},
\newblock \bibinfo{title}{Modeling the microrheology of inhomogeneous media},
\newblock \bibinfo{journal}{Journal of Non-Newtonian Fluid Mechanics}
  \bibinfo{volume}{193} (\bibinfo{year}{2013}) \bibinfo{pages}{21 -- 27}.
  \bibinfo{note}{Viscoplastic Fluids: From Theory to Application}.
\bibitem[{Oppong et~al.(2006)Oppong, Rubatat, Frisken, Bailey, and
  de~Bruyn}]{debruyn1}
\bibinfo{author}{F.~K. Oppong}, \bibinfo{author}{L.~Rubatat},
  \bibinfo{author}{B.~J. Frisken}, \bibinfo{author}{A.~E. Bailey},
  \bibinfo{author}{J.~R. de~Bruyn},
\newblock \bibinfo{title}{Microrheology and structure of a yield-stress polymer
  gel},
\newblock \bibinfo{journal}{Phys. Rev. E} \bibinfo{volume}{73}
  (\bibinfo{year}{2006}) \bibinfo{pages}{041405}.
\bibitem[{Oppong and de~Bruyn(2007)}]{debruyn2}
\bibinfo{author}{F.~K. Oppong}, \bibinfo{author}{J.~R. de~Bruyn},
\newblock \bibinfo{title}{Diffusion of microscopic tracer particles in a
  yield-stress fluid},
\newblock \bibinfo{journal}{J. Non-Newtonian Fluid Mech.} \bibinfo{volume}{142}
  (\bibinfo{year}{2007}) \bibinfo{pages}{104--111}.
\bibitem[{Ising(1925)}]{ising}
\bibinfo{author}{E.~Ising},
\newblock \bibinfo{title}{Beitrag zur theorie des ferromagnetismus},
\newblock \bibinfo{journal}{Zeitschrift f\"{u}r Physik} \bibinfo{volume}{31}
  (\bibinfo{year}{1925}) \bibinfo{pages}{253--258}.
\bibitem[{Stanley(1987)}]{stanley}
\bibinfo{author}{E.~H. Stanley}, \bibinfo{title}{Phase transitions and critical
  phenomena}, \bibinfo{publisher}{Oxford University Press},
  \bibinfo{year}{1987}.
\bibitem[{Landau and Lisfshits(1980)}]{landaust}
\bibinfo{author}{L.~D. Landau}, \bibinfo{author}{E.~M. Lisfshits},
  \bibinfo{title}{Statistical {P}hysics, Part 1: Volume 5 (Course of
  Theoretical Physics, Volume 5)}, \bibinfo{publisher}{Butterworth-Heinemann},
  \bibinfo{edition}{third edition} edition, \bibinfo{year}{1980}.
\bibitem[{Hofschuster and Kr{\"a}mer(2003)}]{HofschusterK03}
\bibinfo{author}{W.~Hofschuster}, \bibinfo{author}{W.~Kr{\"a}mer},
\newblock \bibinfo{title}{C-xsc 2.0: A c++ library for extended scientific
  computing},
\newblock in: \bibinfo{booktitle}{Numerical Software with Result Verification},
  pp. \bibinfo{pages}{15--35}.
\bibitem[{Souli{\`e}s et~al.(2013)Souli{\`e}s, Pruvost, Legrand, Castelain, and
  Burghelea}]{chlorella}
\bibinfo{author}{A.~Souli{\`e}s}, \bibinfo{author}{J.~Pruvost},
  \bibinfo{author}{J.~Legrand}, \bibinfo{author}{C.~Castelain},
  \bibinfo{author}{T.~I. Burghelea},
\newblock \bibinfo{title}{Rheological properties of suspensions of the green
  microalga chlorella vulgaris at various volume fractions},
\newblock \bibinfo{journal}{Rheologica Acta} \bibinfo{volume}{52}
  (\bibinfo{year}{2013}) \bibinfo{pages}{589--605}.
\bibitem[{Stauffer et~al.(1982)Stauffer, Coniglio, and Adam}]{Stauffer1982}
\bibinfo{author}{D.~Stauffer}, \bibinfo{author}{A.~Coniglio},
  \bibinfo{author}{M.~Adam},
\newblock \bibinfo{title}{Gelation and critical phenomena},
\newblock \bibinfo{journal}{Advances in Polymer Science} \bibinfo{volume}{44}
  (\bibinfo{year}{1982}) \bibinfo{pages}{103--158}.

\end{thebibliography}

\end{document}